\documentclass{aa}  

\usepackage{graphicx}
\usepackage{txfonts}
\usepackage{natbib}
\usepackage{lipsum}
\usepackage{booktabs}
\usepackage{subcaption}         
                                
\usepackage{lscape}             
                                
\usepackage{placeins}

\begin{document} 

   \title{The optical colors of hidden spiral-like features in nine Virgo early--type dwarfs}

   \author{Mauricio Gracia\inst{1}\corrauth{mgracia@usm.cl}
    \and Rory Smith\inst{1,2}\email{rory.smith@usm.cl}
    \and Anna Pasquali\inst{3}\email{pasquali@ari.uni-heidelberg.de}
    \and Laura Ferrarese\inst{4}\email{Laura.Ferrarese@nrc-cnrc.gc.ca} 
    \and Patrick C\^ot\'e\inst{4}\email{Patrick.Cote@nrc-cnrc.gc.ca} 
    \and Hyowon Kim\inst{5}\email{hwkim1011@gmail.com}
    \and Paula Calder\'on--Castillo\inst{1}\email{pau.astro.cc@gmail.com}
    \and Kirsi Pohl\inst{1}\email{kpohl@usm.cl}
   }
   \institute{Departamento de F\'isica, Universidad T\'ecnica Federico Santa Mar\'ia, Avenida España 1680, Valpara\'iso, Chile.
    \and Millennium Nucleus for Galaxies (MINGAL), Valpara\'iso, Chile.
    \and Astronomisches Rechen-Institut, Zentrum f\"ur Astronomie der Universit\"at Heidelberg, M\"onchhofstr. 12-14, D-69120 Heidelberg, Germany.
    \and National Research Council of Canada, Herzberg Astronomy \& Astrophysics Research Centre, Victoria, BC, V9E 2E7, Canada.
    \and Korea Astronomy and Space Science Institute, Daejeon 34055, Republic of Korea
    }
   \date{Received June 16, 2026; accepted October 6, 2026}

\abstract
{Dwarf early--type galaxies (ETGs) are abundant in nearby clusters and provide a key laboratory for studying environmentally driven galaxy transformation. Although dwarf ETGs were once considered structurally simple, deep imaging has revealed embedded spiral-like features (e.g., bars and spiral arms) in some of them, contributing only a small fraction of the total galaxy light. The origin of these features at dwarf scales remains uncertain.}
{As a pilot study, we investigate the nature of spiral-like features in nine Virgo dwarf ETGs with clear embedded arms. Our goal is to use optical colors to test whether these features differ from the surrounding diffuse component, which would constrain their formation pathways. In doing so we establish the methodology on a well-characterized sample, ahead of a future analysis of the much larger NGVS dwarf sample.}
{Using deep \(g\)- and \(i\)-band images from the NGVS, we apply a residual-based image analysis method to separate excess light associated with the spiral-like component from the diffuse host. The residual images are used to define the arm and interarm regions, enabling direct measurements of light fractions and \(g-i\) colors for the spiral-like feature and the diffuse component in which they are embedded.}
{The spiral-like features contribute only a few percent of the total galaxy light (approximately \( 1.5\%\!-\!5.5\%\) in \(i\) and \( 1.7\%\!-\!3.7\%\) in \(g\)). We find a diverse range of stellar colors across the sample: four galaxies host spiral-like features that are redder than their interarm regions, four show no robust color difference between the arm and interarm regions, and one exhibits a distinctly bluer spiral-like component. The latter is consistent with a recently quenched transition object, while the redder and color-uniform features are consistent with an older or more metal-rich stellar population.}
{Spiral-like features in Virgo dwarf ETGs exhibit a surprising diversity in their photometric properties, implying multiple evolutionary channels are at work rather than a single one. This pilot analysis also demonstrates the readiness for the much larger NGVS sample of dwarfs with potential spiral-like features.}


    \maketitle
    \nolinenumbers

\section{Introduction}\label{sec:intro}

In the local Universe, small, low-mass dwarf galaxies represent the most numerous class of galaxies \citep{1985AJ.....90.1681B,1989AJ.....98..367F}. Morphologically, dwarfs are commonly divided into two broad categories based on their global colors and level of star formation activity \citep{1984AJ.....89..919S}: dwarf early--type galaxies (dETGs) and dwarf late-type galaxies (dLTGs). Both are characterized by low luminosities with the bright end being $M_B \approx -18$ \citep{1994A&ARv...6...67F}  and low stellar masses ($M_\star \lesssim 10^{9-9.5}\,{\rm M_\odot}$). 

Dwarf LTGs exhibit blue colors, ongoing star formation, and substantial reservoirs of gas and dust. They typically show irregular morphologies ( e.g., \citealt{1989ApJ...347..760C,1989ApJ...347..875S,2003ApJ...591..167C,2005NewAR..49..453S,2015ApJS..217...27A}) or spiral-like features containing spiral arms or bars (e.g., \citealt{2001ApSSS.277..231G}) and they are preferentially located in the lower-density outskirts of galaxy clusters, where dwarfs tend to rotate more strongly than those in the inner cluster. In contrast, dwarf ETGs are largely quenched systems with little gas or dust, resulting in their globally red optical colors and generally smooth morphology.  Historically divided into dwarf ellipticals \citep{1994A&ARv...6...67F,2003AJ....125.2936G}, dwarf lenticulars \citep{1984AJ.....89..919S,1991A&A...252...27B}, and dwarf spheroidals \citep{2001AJ....122.3092H}, dwarf ETGs dominate the galaxy population in dense environments \citep{1985AJ.....90.1681B,2025AJ....170..363K}, while their occurrence in the field is very rare, with a probability below $\sim0.06\%$ \citep{2012ApJ...757...85G}.

Dwarf ETGs, despite their seemingly simple appearance, have been discovered to harbor a range of intricate structural and stellar population complexities. Many are not well described by a single Sersic profile \citep{1968adga.book.....S}, exhibiting multiple components \citep{2012ApJ...745L..24J,2014ApJ...786..105J,2021A&A...647A.100S}; moreover, spiral arms and bars have been detected in several cluster ETGs \citep{2001ASPC..230..239J,2002A&A...391..823B,2003A&A...400..119D,2003AJ....125.2936G,2006ApJS..164..334F,2006AJ....132..497L,2014MNRAS.443.3381P,2019A&A...625A.143V,2022AJ....164...18M}. Color imaging has revealed that some of these galaxies contain blue central regions indicative of recent or ongoing star formation, even though the class as a whole resides on the red sequence of the color–magnitude diagram \citep{1973ApJ...182..671H,1984A&A...139L...9V,1993A&A...271...51P,2006AJ....132.2432L,2014MNRAS.445..630P,2017A&A...606A.135U,2019A&A...625A..94H}. These "blue-cored" dwarf ETGs are preferentially located in the lower-density outskirts of galaxy clusters where dwarfs tend to be rotating more strongly than dwarf in the inner cluster \citep{2009AN....330.1043L,2009ApJ...707L..17T}. Almost all ETGs have negative metallicity gradients \citep{2017A&A...606A.135U,2017MNRAS.470..815S, 2023MNRAS.525.4329B}. Some of these steep gradients are linked to past dwarf-dwarf merger events \citep{2022A&A...665A..92T}, while others are associated with the presence of shells in certain cluster members \citep{2017ApJ...834...66P}. Evidence of morphological and stellar population diversity implies that dwarf ETGs do not originate from a single progenitor, but instead emerge through multiple evolutionary pathways (see \citealp{2009AN....330.1043L}).

In the literature, these pathways are generally grouped into two broad classes: "nature" and "nurture" scenarios. In the nature scenario, the properties of dwarf ETGs reflect their intrinsic formation histories, such as early collapse, gradual mass assembly, or minor mergers that can heat or reshape their stellar components. In nurture, environmental mechanisms acting after infall into a group or cluster are responsible for quenching star formation and transforming LTG progenitors into ETGs. Several processes can contribute to such transformations, including starvation or strangulation, which halts the supply of fresh gas as the galaxy falls into high-density regions \citep{1980ApJ...237..692L,2008ApJ...674..742B}; ram-pressure stripping by the hot intracluster medium, which efficiently removes cold gas \citep{1972ApJ...176....1G,2004AJ....128.2797V,2010ApJ...724L.171D}; and tidal harassment from high-speed encounters or the global cluster potential, which primarily leads to dynamical heating and thickening of stellar disks \citep{2010MNRAS.405.1723S,2015MNRAS.454.2502S} and can perturb internal structures \citep{1998ApJ...495..139M,1999MNRAS.304..465M,2022AJ....164...18M}.

Early systematic efforts, most notably by \citet{2006AJ....132..497L}, used an iterative technique similar to unsharp masking on SDSS imaging to identify and study spiral-like substructures such as bars and spiral arms in Virgo dwarf ETGs. These studies first established that a nonnegligible fraction of Virgo dwarf galaxies exhibit embedded spiral-like components, and that residual-based decomposition can delineate where those components lie within a galaxy. While their contribution to the total galaxy light is typically modest (between $5\%$--$15\%$), it can rise to nearly half in some dwarf ETGs.

A more refined approach was introduced by \citet{2021AJ....161..268M}, who formalized and extended these residual-based techniques into an iterative procedure,  referred to as the "residual method", applied to nine Virgo dwarf ETGs with prior evidence of embedded substructures in SDSS DR4 imaging \citep{2006AJ....132..497L}. Using deep white-light wide field imager (WFI) imaging, they find that the hidden structures contribute between $2.2\%$ and $6.4\%$ of the total galaxy light within two effective radii ($R_{\mathrm{e}}$). However, the use of a single broad white filter meant that little information could be recovered on the stellar populations of the hidden spiral-like features, leaving open the question of whether they differ from the main body of the galaxy.

With deep multiwavelength photometry it becomes possible to search for color differences between the spiral-like features and their host galaxy. The presence of such differences would demonstrate that the two components host distinct stellar populations, and would already begin to place constraints on the origin of the features. The well-known age--metallicity degeneracy, however, means that a color difference alone cannot reveal whether it arises from a difference in age, in metallicity, or in both, and answering that question will ultimately require spectroscopy. Deep multiwavelength imaging is nevertheless of considerable value to spectroscopic studies in its own right. Dwarf galaxies are intrinsically faint and of low surface brightness, which already makes spectroscopy expensive, and the spiral-like features of interest typically contribute only a few percent of the total galaxy light, so that they are swamped by the light of the main stellar body. Deep imaging combined with the iterative residual method can identify the objects of interest, pinpoint the exact locations of the spiral-like features, and thereby provide crucial constraints on where to place spectroscopic slits or which IFU spaxels to stack. It also measures the fraction of light contained in the arms, which allows the feasibility of a spectroscopic study to be assessed and integration times to be estimated, while the broadband colors offer an educated guess as to whether emission lines are likely to be present. We therefore regard deep multiwavelength imaging, combined with the residual method, as a crucial prerequisite,or at the very least a highly valuable preliminary step, toward successful spectroscopic studies of these objects.

The majority of dwarf ETGs with spiral-like features that have been found and studied belong to the Virgo cluster, owing to its proximity \citep[approximately $16.5\,$Mpc;][]{2009ApJ...694..556B} and to the fact that it has been extensively cataloged and surveyed; see, the Virgo Cluster Catalog \citep[VCC;][]{1985AJ.....90.1681B}; the Advanced Camera for Surveys Virgo Cluster Survey \citep{2004ApJS..153..223C}; The Next Generation Virgo Cluster Survey \citep[NGVS;][]{2012ApJS..200....4F}. However, the origin of these spiral-like features is still an object of scientific debate. It is often suggested that dwarf ETGs are the remnants of infalling dwarf LTGs that were quenched by environmental transformation \citep{1998ApJ...495..139M,2014MNRAS.443.3381P}. Another possibility, not necessarily mutually exclusive, is that the environment may trigger the formation of these features in dwarf ETGs; \citet{2021ApJ...912..149S} used high-resolution simulations of cluster harassment and find that spiral-like features, like bars and spiral arms, can arise when a rotationally supported, dynamically cold stellar disk is embedded within a thicker, hotter component. Tidal interactions with the cluster potential can then trigger the formation of spiral arms and bar features, even in the absence of star formation.

In this work we present a pilot color analysis of hidden spiral-like features in Virgo dwarf ETGs based on deep multiband NGVS imaging \citep{2012ApJS..200....4F}, providing optical color information at a depth comparable to \citet{2021AJ....161..268M}. We deliberately revisit the same nine galaxies analyzed by \citet{2021AJ....161..268M} with a primary goal of measuring the optical color (\(g-i\)) of hidden spiral-like features. With this calibrated technique in hand, we use the residual method to isolate the location of spiral-like features (arms, bars, and disks), measure their light fractions and \(g-i\) colors and compare them to those of the surrounding diffuse host. The aim is to test whether these embedded features exhibit systematic color differences relative to the main body. Such a test bears directly on the nature and nurture scenarios outlined above. If the features are the recently transformed descendants  of infalling dwarf LTGs \citep{1998ApJ...495..139M,2014MNRAS.443.3381P}, the arms which hosted the bulk of  the star formation prior to quenching should appear bluer than the diffuse body \citep{2006AJ....132..497L,2014MNRAS.445..630P,2017A&A...606A.135U}. However, if the quenching occurred sufficiently long ago, the blue arms may fade to red colors, becoming similar in color to the diffuse component. Alternatively, \citet{2021ApJ...912..149S} proposed that a high rotation support/dynamically cold disk can be tidally triggered to form bars and spiral features without triggering the dynamically hotter surrounding disk. In this scenario, a thin cold disk might hide within the dominating thick disk, only revealing itself after a strong tidal shock such as a cluster pericenter passage. If the thin disk is also old and red, having hidden for a long time, the colors of the spiral features that are generated might be indistinguishable from the diffuse component. Finally, if the spiral-like feature is more metal-rich than its surroundings as reported for the central regions of some Virgo dwarf ETGs \citep{2017A&A...606A.135U,2022MNRAS.515.4622B}, the features should appear redder at comparable age. Broadband colors alone cannot break the age--metallicity degeneracy (Sect.~\ref{sec:Discussion}), but they can establish which regime each galaxy occupies.

This paper is organized as follows. Sect.~\ref{sec:Data_and_Observations} describes the sample selection, the NGVS data, the preprocessing steps and the global properties of the sample. Sect.~\ref{sec:Residual_Method} presents the residual method \citep{2021AJ....161..268M}, its multiband extension \citep{2022AJ....164...18M}, the revised jump-out criterion introduced in this work and the resulting residual light fractions. Sect.~\ref{sec:Results} reports the \(g-i\) color measurements of the spiral-like features and the accompanying significance analysis. In Sect.~\ref{sec:Discussion} we interpret the results in the context of formation channels and environmental processing and Sect.~\ref{sec:Summary} summarizes our conclusions.

\section{Data and observations}\label{sec:Data_and_Observations}

\subsection{Sample selection}\label{subsec:sample_selection}

This work is based on $g$- and $i$-band imaging from NGVS \citep{2012ApJS..200....4F}, the deepest wide-field optical survey of the Virgo cluster to date. The NGVS covers $104~\mathrm{deg}^{2}$ around the two main subclusters centered on M87 and M49, reaching a $2\sigma$ limiting surface brightness of $\mu_{g}\simeq29\;\mathrm{mag\,arcsec^{-2}}$ for extended emission. Imaging was obtained with MegaCam ($0\farcs187\;\mathrm{pixel^{-1}}$) on the 3.6\,m Canada-France-Hawaii Telescope and processed through the \textsc{Elixir-LSB} and \textsc{MegaPipe} pipelines \citep{2012ApJS..200....4F}, which perform bias subtraction, flat-fielding, scattered-light removal, astrometric calibration, and image stacking. Typical seeing in the final stacked images is $0\farcs6$--$0\farcs9$, corresponding to physical scales of $\sim$50--70\,pc at the adopted Virgo distance of 16.5\,Mpc \citep{2009ApJ...694..556B}. Among the four NGVS broadband filters ($u^{*}$, $g$, $i$, $z$), we focus on the $g$ and $i$ bands, which reach the deepest surface-brightness limits while the $(g{-}i)$ color still brackets the bulk of the optical stellar light. Full details on the survey, including observing, data processing and sky subtraction, are available in \citet{2012ApJS..200....4F}.

\begin{table*}[h]

\caption{Photometric and morphological properties of the dwarf ETG sample.}
\label{tab:photometric_properties}
\centering
$$
\begin{array}{ccccccc}
\hline
\noalign{\smallskip}
\text{VCC} & \text{NGVS} & \text{PSF-FWHM} & M_i & g{-}i  & \text{Galaxy Type} \\

              &    & \text{[arcsec]} &  \text{[mag]} & \text{[mag]}  &  \text{from NGVS}\\

(1)             &        (2)            &          (3)            & (4)  & (5)  & (6)  \\
\noalign{\smallskip}
\hline
\noalign{\smallskip}
\text{0216} & \text{0259}  & 0.95  & -16.98 \pm 0.15 & 0.93  & \text{ES(n|di:|sp:|du)-I(c)}      \\
\text{0308} & \text{0356}  & 0.88  & -18.17 \pm 0.15 & 0.84  & \text{ES(n|di|sp)-Q}              \\
\text{0490} & \text{0605}  & 0.82  & -18.35 \pm 0.15 & 0.74  & \text{ES(n|di|b|sp|du:)-Q}        \\
\text{0523} & \text{0649}  & 0.93  & -18.76 \pm 0.15 & 0.84  & \text{ES(n|di:|b|sp:|sh|st:)-I(c)}\\
\text{0856} & \text{1219}  & 0.88  & -17.92 \pm 0.15 & 0.87  & \text{ES(n|di|sp)-I(g)}           \\
\text{0940} & \text{1387}  & 1.07  & -17.47 \pm 0.15 & 0.97  & \text{ES(n|di|b|sp)-Q}            \\
\text{1010} & \text{1517}  & 1.02  & -18.67 \pm 0.15 & 0.97  & \text{ES(n|di|b|sh:|sp)-Q}        \\
\text{1695} & \text{2911}  & 0.85  & -17.78 \pm 0.15 & 0.75  & \text{ES(di|b|sp|st)-Q}           \\
\text{1896} & \text{3266}  & 0.98  & -17.20 \pm 0.15 & 0.95  & \text{ES(n|di|b|sp|)-Q }          \\

\noalign{\smallskip}
\hline
\end{array}
$$

\tablefoot{ Column (1): galaxy identifier from \citet{1985AJ.....90.1681B}.  Column (2): NGVS catalog identifier.  Column (3): final spatial resolution after convolution with the filter PSF.  Column (4): absolute \(i\)-band magnitude assuming a Virgo Cluster distance modulus of \(31.09 \pm 0.15\) mag \citep{2009ApJ...694..556B}.  Column (5): integrated \(g-i\) color.  Column (6): morphological and star-formation classification from the NGVS catalog.  The primary morphology codes are: E = elliptical, ES = elliptical/spiral transition, S = spiral, SI = spiral/irregular transition, I = irregular, and EI = elliptical/irregular transition.  Morphological subcodes are: sp = spiral arms, di = disk, b = bar, sh = shells, du = dust, st = stellar stream, r = ring, fi = filaments, and pm = post-merger remnant.  Star-formation classes are: Q = quiescent, I = intermediate, and A = active.  Star-formation subcodes are: g = global, c = core, f = filamentary, p = patchy, e = extended, d = disk star formation, sp = star formation in spiral arms, sh = star formation in shells, and r = star formation in rings.  A colon ``:'' indicates an uncertain classification. All photometric measurements are given in the AB magnitude system and corrected for Galactic foreground extinction. }
\end{table*}

All nine galaxies in our sample are classified in the NGVS visual-morphology catalog as elliptical to spiral transition systems (ES; see Table~\ref{tab:photometric_properties}, column~6). This class denotes a smooth, red, dETG-like body that shows hints of spiral-like features, such as spiral arms, bars, or disks. They are therefore drawn from the same parent population as a substantially larger set of approximately $100$ NGVS galaxies that likely contain spiral-like features \citep{2025AJ....170..363K}. The current sample is small and biased towards systems with already-detected spiral-like features \citep{2006AJ....132..497L,2021AJ....161..268M}. The colors of these features have not previously been measured with deep imaging. Therefore, we consider this work a pilot study that establishes the multiband residual-image technique on a well-characterized subset before extending it to a statistically representative NGVS sample.

\begin{table}[]
\caption{Structural Properties of the Dwarf ETGs Sample.}
\label{tab:structural_properties}
\centering
$$
\begin{array}{lccccc}
\hline
\noalign{\smallskip}
\text{VCC} & \text{Effective radius} & \multicolumn{2}{c}{\text{Ellipticity}}  & \multicolumn{2}{c}{\text{Position angle}}\\

 & 1\,R_{\mathrm{e}} & 1\,R_{\mathrm{e}}  & 2\,R_{\mathrm{e}} & 1\,R_{\mathrm{e}} & 2\,R_{\mathrm{e}} \\

              & \text{[arcsec]} &   &   & \text{[deg]} & \text{[deg]} \\

(1)             &        (2)            &          (3)            & (4)  & (5)  & (6)  \\
\noalign{\smallskip}
\hline
\noalign{\smallskip}
\text{0216} & 13.2 & 0.3  & 0.3  & 28   & 25   \\
\text{0308} & 19.4 & 0.1  & 0.1  & 76   & 76   \\
\text{0490} & 31.2 & 0.2  & 0.2  & 308  & 308  \\
\text{0523} & 25.8 & 0.3  & 0.3  & 318  & 318  \\
\text{0856} & 16.0 & 0.1  & 0.1  & 81   & 85   \\
\text{0940} & 21.3 & 0.2  & 0.2  & 17   & 17   \\
\text{1010} & 21.3 & 0.4  & 0.4  & 356  & 356  \\
\text{1695} & 23.5 & 0.2  & 0.2  & 83   & 83   \\
\text{1896} & 17.6 & 0.3  & 0.1  & 23   & 284  \\
\noalign{\smallskip}
\hline
\end{array}
$$
\tablefoot{Column (1): name of dwarf galaxy as labeled in \citet{1985AJ.....90.1681B}. Column (2): effective semimajor axis length of the elliptical isophote in the \(i\)-band. Columns (3) and (4): ellipticity at one and two effective radii, respectively. The ellipticity is defined as $e = 1-ba^{-1}$, where b and a are the semiminor and semimajor axis length of the isophote at that radius. Columns (5) and (6): position angle at one and two effective radii, respectively. The position angle is measured counterclockwise from the positive y-axis (north toward east of the image).}
\end{table}

\subsection{Data preparation}\label{subsec:data_preparation}
  
Prior to applying the residual method, we performed a set of preprocessing steps to ensure reliable photometric measurements of the faint embedded substructures.
 
To minimize contamination from interloping foreground stars and background galaxies, we used a bad pixel mask (BPM), described in \citet[Sect.~5.1.1]{2020ApJ...890..128F}. We matched the point-spread functions (PSFs) of the $g-$ and $i-$band images using a cross-convolution scheme, convolving each band with the PSF model of the other. The advantage of this approach is operational: it equalizes the effective resolution of the two bands without requiring an analytical or fitted target PSF, removing one possible source of systematic error in the color maps. The trade-off is that both images are degraded to an effective resolution slightly worse than that of either original band.

We then fitted galaxy isophotes to the PSF-matched images with the \textsc{IRAF} \texttt{ELLIPSE} task, allowing the center position, ellipticity, and position angle to vary freely with the semimajor axis. The BPM was supplied to \texttt{ELLIPSE} so that contaminated pixels are excluded from the fitting. We defined the galaxy boundary as the outermost isophote whose mean intensity lies at least $2\sigma$ above the background noise level. We estimated the background using \texttt{SExtractor} \citep{1996A&AS..117..393B} \texttt{BACKGROUND} and \texttt{BACKGROUND RMS} images, with a mesh size (\texttt{BACK\_SIZE}) of 64 pixels$^2$. 

All magnitudes and surface brightnesses reported in this work are in the AB system and have been corrected for Galactic foreground extinction using the reddening maps of \citet{2011ApJ...737..103S}, with extinction coefficients $A_{g}/E(B{-}V)=3.560$ and $A_{i}/E(B{-}V)=1.813$ derived from the \citet{1999PASP..111...63F} extinction law via the York Extinction Solver.

\subsection{Structural parameters}\label{subsec:observations}

From the PSF-matched isophotal profiles we derived the structural and photometric properties of the sample. we measured total flux within the outermost valid isophote (defined in Sect.~\ref{subsec:data_preparation}) and  defined the effective radius $R_{\mathrm{e}}$ as the semimajor axis of the isophote enclosing half of this total flux. The resulting photometric properties are listed in columns 4 and 5 of Table~\ref{tab:photometric_properties}. We note that the exact magnitudes can differ slightly from those reported by other studies using NGVS data, owing to differences in the adopted photometric measurement choices. The galaxies are intrinsically faint, with absolute magnitudes in the range $-18.76 \leq M_{i} \leq -16.98$\ and relatively uniform average colors of $0.74 \leq (g{-}i) \leq 0.97$\,mag, consistent with a sample of low-luminosity, moderately red dwarf galaxies. 
 
The galaxy center was defined as the centroid of the isophote at a semimajor axis of $1.5 \times \mathrm{PSF}_{\mathrm{FWHM}}$, which prevents any effect of the PSF when determining the central coordinates. The area enclosed by this isophote was masked and added to the BPM to alleviate contamination from unresolved nuclear light in subsequent analyses; this also prevents fitting artifacts that the \texttt{ELLIPSE} task can produce in the steep central gradient. The masked region represents a negligible fraction of the effective radius, so no significant galaxy light is lost. The effective radius, ellipticity, and position angle measured at \(1\,R_{\mathrm{e}}\) and \(2\,R_{\mathrm{e}}\) are presented in Table~\ref{tab:structural_properties}. These measurements define the reference geometry used to configure the residual method described in Sect.~\ref{sec:Residual_Method}.

\section{The residual method}\label{sec:Residual_Method}

The residual method, introduced by \citet{2021AJ....161..268M}, is an iterative image-analysis procedure designed to separate a galaxy image into two distinct components: (i)~a galaxy model image that contains the dominant, smooth (axisymmetric) light of the diffuse stellar body, and (ii)~a galaxy residual image that contains the secondary, nonaxisymmetric light associated with embedded  substructures such as spiral arms and bars.

The procedure works as follows. The galaxy image is first convolved with an elliptical Gaussian kernel (using the \texttt{IRAF.GAUSS} task) whose shape matches the global geometry of the galaxy at two effective radii. The kernel size is tuned to blur out the spiral-like  features while preserving the overall light distribution. The smoothed image is then fitted with concentric, free--ellipticity isophotes (\texttt{IRAF.ELLIPSE}), yielding a geometric description of the radial light profile that is minimally influenced by the substructures. These isophotes are subsequently imposed on the original (unsmoothed) galaxy image in photometry-only mode to recover the true average brightness along each contour. From these measurements a two-dimensional model of the diffuse component is built with \texttt{IRAF.BMODEL}, and subtracting this model from the original image produces the galaxy residual image, which ideally isolates the spiral-like component light.

In practice, however,  some light from the spiral-like features inventively contaminates the model after a single-pass, causing the residual image to suffer from over-subtraction: the interarm regions are dominated by negative pixel values. To correct for this, the positive residual light (associated with the spiral-like features) is subtracted from the original galaxy image, producing an updated image in which the contribution of the spiral-like features is reduced. This updated image then re-enters the procedure, and the loop of smoothing, fitting, modeling, and subtracting is repeated until a convergence criterion is met. The iterations stop at the jump-out iteration (JOI), defined as the first iteration at which the median pixel value of the interarm region in the residual image reaches a value $\geq 0$, indicating that over-subtraction has been eliminated without introducing under-subtraction. A detailed, step-by-step description is given in Sect.~3 of \citet{2021AJ....161..268M}.

We note that the residual method differs from the unsharp masking, which is traditionally used to reveal substructure in dETGs \citep[e.g.,][]{2006MNRAS.370..477L,2006AJ....132.2432L}. In unsharp masking, the image is divided by, or has subtracted from it, a smoothed copy of itself using a kernel of fixed size; the technique is fast and effective at revealing where a feature is, but the amplitude of the enhanced structure depends on the kernel size in a way that has no direct photometric interpretation, and the smoothed copy is not a model of the diffuse component. The residual method instead builds an explicit axisymmetric model of the host from free-ellipticity isophote fit, the original spatial resolution is preserved,and the residual image is expressed in physical flux units and can be integrated to yield light fractions and, in the multiband extension used here, colors. The iterative loop additionally corrects for contamination of the model by the substructure light itself, the dominant systematic error of any single-pass subtraction. A more detailed comparison is given in \citet{2021AJ....161..268M}.

It is important to note that, throughout this work, the spiral-like feature refers to the nonaxisymmetric, excess light contained in the embedded substructures. It does not correspond to the total luminosity of a physically embedded spiral, which may also contribute a smooth component that is inseparable from the diffuse body of the galaxy \citep{2021AJ....161..268M}.

\subsection{Method setup}\label{sec:method_setup}

The residual method has two free parameters common to both single-filter and multiband applications, plus one additional parameter specific to the multiband setup. The first parameter is the smoothing kernel size (\texttt{smooth}), which controls the standard deviation of the elliptical Gaussian kernel applied in the convolution step. Following \citet{2021AJ....161..268M}, this size should match approximately half the average width of the spiral-like features, so that the substructures are effectively blurred out without significantly altering the overall galaxy geometry. For our sample we expressed this quantity in units of the PSF standard deviation ($\sigma_{\mathrm{PSF}}$) and explored a range from $6\,\sigma_{\mathrm{PSF}}$ to $12\,\sigma_{\mathrm{PSF}}$, stepped in increments of $1\,\sigma_{\mathrm{PSF}}$ (seven values in total). As recommended by \citet{2021AJ....161..268M}, the kernel size is halved after the first iteration, since the majority of the substructure light is removed in the initial pass and further smoothing should be kept to the minimum necessary to preserve the geometry of the diffuse component.

The second parameter is the isophotal step size (\texttt{step-size}), which sets the geometric growth rate of the semimajor axis between consecutive isophotes during the \texttt{ELLIPSE} fitting. Smaller steps provide finer radial sampling of the light profile but require more data points per isophote; larger steps are more robust in low S/N regions but may miss subtle radial variations. We adopted a geometric growth rate ranging from $4\%$ to $10\%$, also in increments of one percentage point (seven values). Together, the smoothing and step-size grids yield a total of $7 \times 7 = 49$ parameter combinations per galaxy.

The third parameter, applicable only in the multiband setup, is the reference filter (\texttt{main-filter}). In a multiband analysis, the isophotal geometry must be derived in a single filter and then imposed on all other bands, so that the spatial decomposition is consistent across wavelengths \citep{2022AJ....164...18M}. We adopted the $i$~band as the reference filter for all galaxies because of its greater depth. We first ran the full iterative procedure on the $i$-band images and subsequently imposed the resulting isophotal geometry to the $g$-band images to obtain the corresponding residual images.

\subsection{A revised jump-out criterion}\label{sec:new_joi}

The iterative nature of the residual method enables a progressively cleaner separation between embedded substructures and the diffuse component. In the implementation of \citet{2021AJ....161..268M}, the convergence criterion monitors the median pixel value of an interarm region in the residual image, defined as all pixels that are negative in the first (pre iteration) residual and that lie within the $2\,R_{\mathrm{e}}$ isophote. As the iterations proceed, this region becomes less negative; the loop terminates once its median first reaches a value $\geq 0$, thereby avoiding both over- and under-subtraction.

A limitation of this definition is that pixels that happen to have residual values $\geq 0$ already in the first pass are permanently excluded from the interarm region and implicitly labeled as "arm" pixels, even when they are not associated with genuine spiral-like feature. While this mislabeling is often negligible in brighter systems, the low surface brightness and small angular size of the features in our targets motivate a stricter definition.

We therefore retain the JOI framework of \citet{2021AJ....161..268M} but focus on accurately identifying where the substructures are. Starting from an initial \textsc{SExtractor} segmentation map of the residual image, we applied a small set of morphological operations (erosion, dilation, and connected-component filtering; details in Appendix~\ref{apx:new_mask_steps}) designed to remove noise-driven detections and isolated mislabeled pixels while preserving contiguous spiral-arm footprints. The resulting arm measuring mask (AMM) provides a faithful delineation of the substructure regions. Its complement defines the interarm measuring mask (IAMM), which encompasses all nonarm pixels within the galaxy: a mixture of incompletely subtracted diffuse-component light and sky-level fluctuations. Because the IAMM now excludes spurious arm detections, it yields a more reliable estimate of the median nonarm level used to monitor convergence during the iterative loop.

\begin{figure}
        \centering
        \includegraphics[width=\hsize]{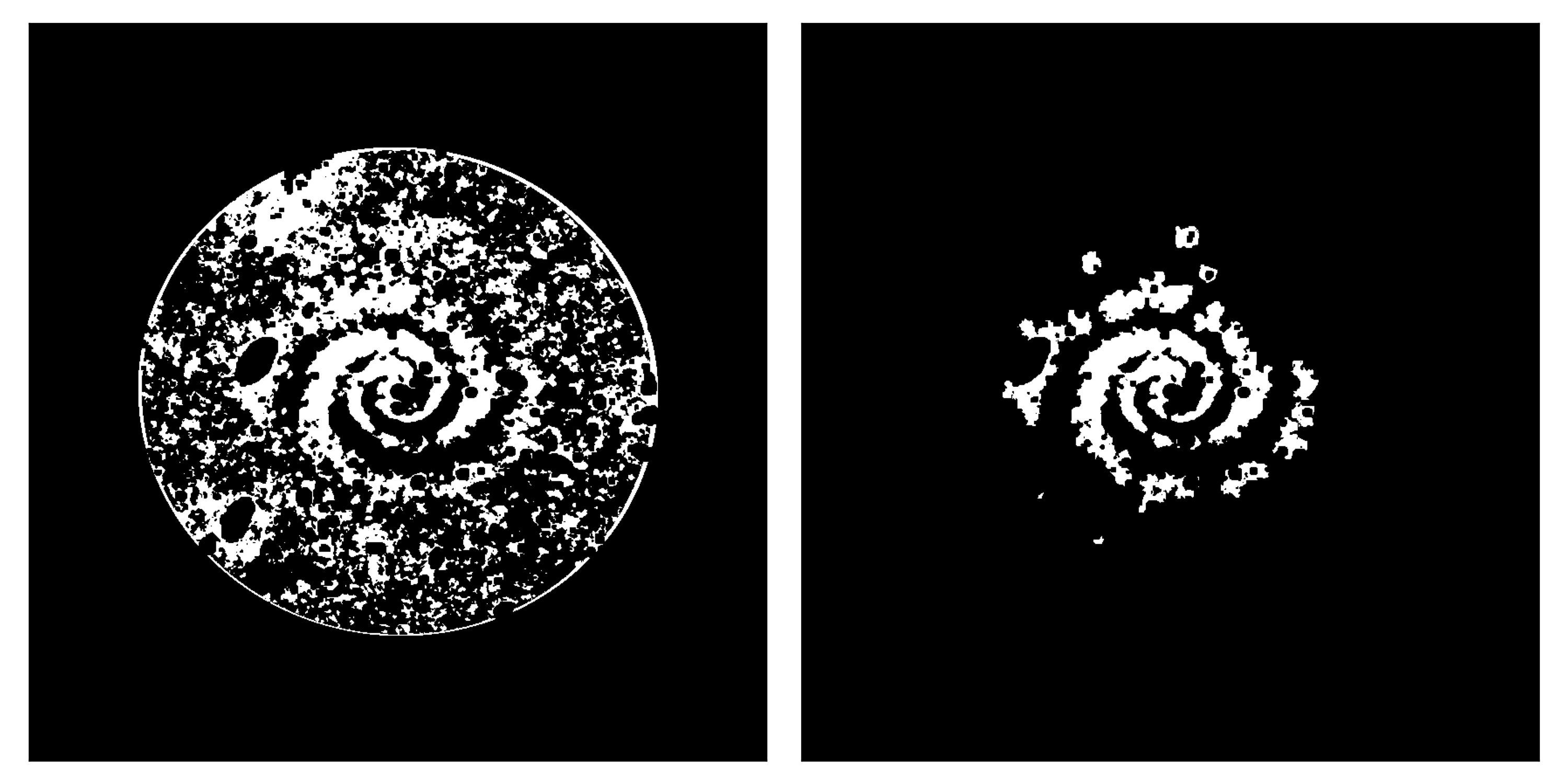}
        \caption{AMMs obtained with the original jump-out criterion (left) and our revised criterion (right), for VCC\,0856. These masks correspond to a smoothed-image kernel of \(10\,\sigma_{\mathrm{PSF}}\) and a step size of \(7\%\). revised mask removes excess outer-region pixels and gives a cleaner delineation of the spiral-arm footprints.}
        \label{fig:mask_comparasion}%
\end{figure}

Figure~\ref{fig:mask_comparasion} compares the original and revised measuring masks for VCC\,0856. The new procedure removes spurious detections in the outer regions while retaining a cleaner and more continuous delineation of the spiral arms. By grounding the arm identification in coherent-structure detection rather than a simple sign-based pixel classification, we reduce the risk of misclassifying nonarm fluctuations as spiral-arm light and improve the robustness of the decomposition for low-contrast features.

The main practical consequence of the revised criterion is that the iterative loop terminates  $\sim$2 iterations early, on average, than in the original implementation, so the final residual images contain less accumulated flux. This does not compromise our analysis for two reasons. First, we use the residual images primarily to define the spatial locations of the spiral-like features (i.e., to construct the AMM and IAMM), rather than to interpret the residual pixel values as a direct measure of the spiral-like feature luminosity. Second, the light fractions measured through the revised masks follow the same  trend as those reported by \citet{2021AJ....161..268M}; this tendency agreement is shown in Fig.~\ref{fig:residual_fraction} and discussed further in Sect.~\ref{sec:residual_results}.

The AMM and IAMM adopted for all subsequent measurements are constructed by applying the morphological cleaning procedure described above to the residual image at the exit iteration. In addition,  we excluded from both masks all pixels flagged in the BPM, which includes masked foreground sources and the central region within $1.5 \times \mathrm{PSF}_{\mathrm{FWHM}}$ (see Sect.~\ref{subsec:observations}). This ensures that the photometric measurements of the arm and interarm regions are not contaminated by light from bright interlopers or by artifacts arising from the PSF-dominated core of the galaxy.

\subsection{Residual results}\label{sec:residual_results}

We applied the residual method to the \(g\)- and \(i\)-band images of nine Virgo dwarf galaxies from the NGVS. For completeness we ran both jump-out criteria: the original definition of \citet{2021AJ....161..268M} (hereafter original JOI) and the modified prescription introduced in Sect.~\ref{sec:new_joi} (new JOI). Figure~\ref{fig:residuals} presents the \(i\)-band residual images obtained with the new JOI; the corresponding \(g\)-band residuals are visually indistinguishable from the \(i\)-band ones, and both JOI variants recover the same structures. For most of the sample the spiral-like features are fully enclosed within the $1\,R_{e}$ ellipse (VCC\,0216, VCC\,0490, VCC\,0523, VCC\,0940, VCC\,1695); those whose arms extend farther (VCC\,0308, VCC\,0856, VCC\,1010, and VCC\,1896) are still contained within the $2\,R_{e}$ isophote. The close agreement with the structures published by \citet{2021AJ....161..268M}, recovered here from only two broadband filters rather than their single broad white-light WFI image, confirms both the robustness of the algorithm and the benefit of the higher-resolution NGVS data.

\begin{figure}[h!]
        \centering
        \includegraphics[width=\hsize]{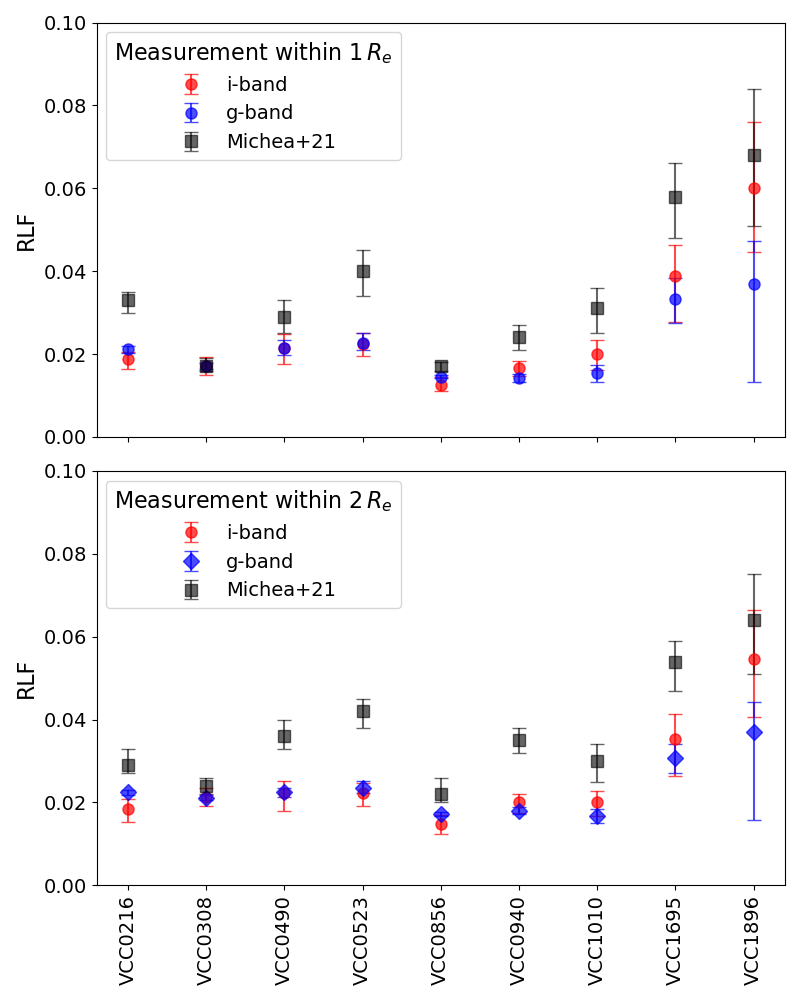}
        \caption{RLF of the dETG sample. Measurements are performed within one (top) and two (bottom) effective radii  for the \(i\) and \(g\) bands, represented in red and blue circles. For comparison, the black squares show the RLF values \citeauthor{2021AJ....161..268M}'s (2021). The symbols show the median RLF of the distribution, and the error bars show the $16^{\text{th}}$ and $84^{\text{th}}$ percentiles. All RLF are corrected for galactic extinction and calculated using the new JOI residual image. Our values are systematically lower than those of \citet{2021AJ....161..268M} owing to the earlier exit of the revised jump-out criterion (Sect.~\ref{sec:new_joi}); the galaxy to galaxy trend is nonetheless preserved.}
        \label{fig:residual_fraction}%
\end{figure}

\begin{figure*}[h!]
        \centering
        \includegraphics[bb=16 5 127 122,width=0.9\hsize,clip]{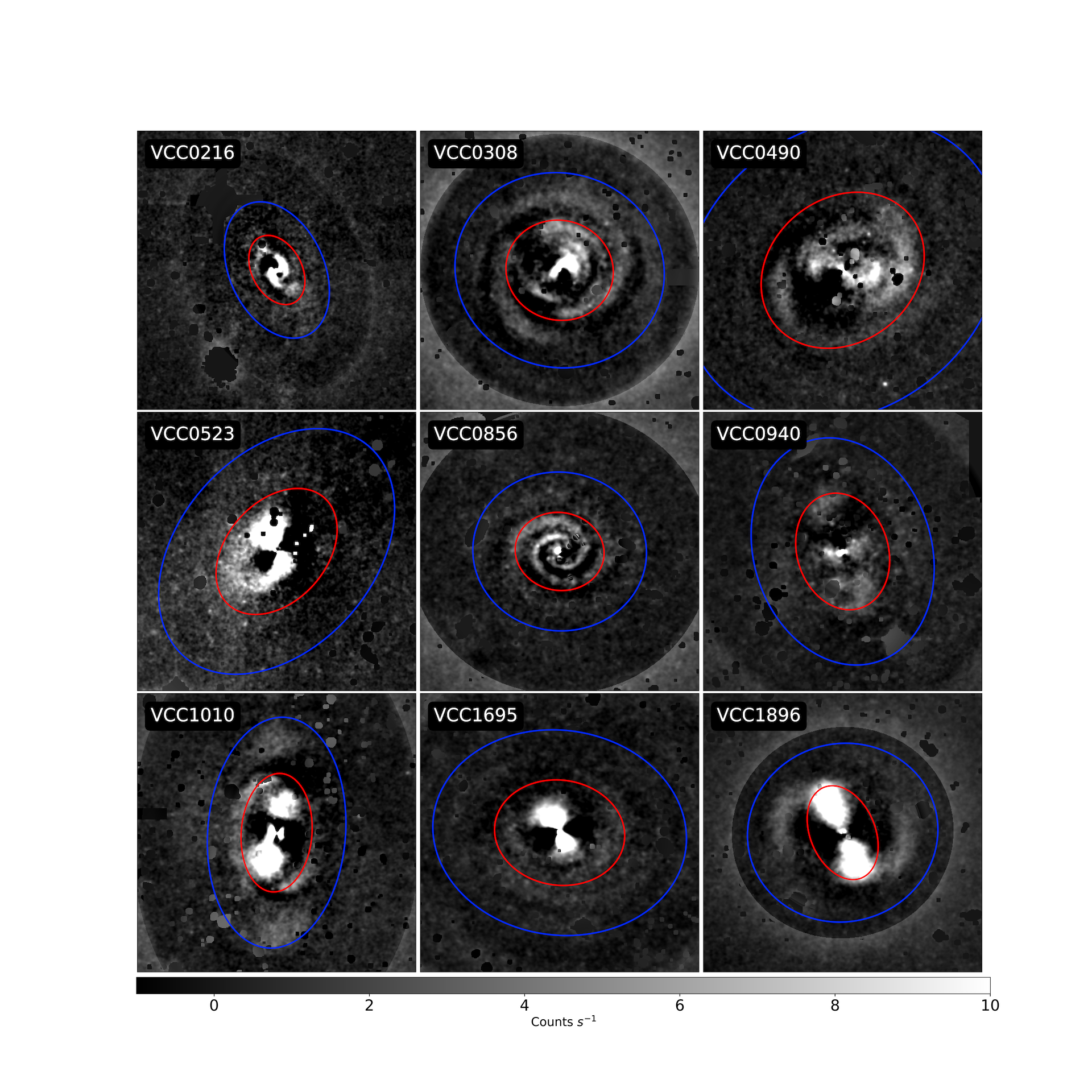}
        \caption{Residual images of the nine Virgo dwarf ETGs for the \(i\) band. The method parameters for these images were a \texttt{smooth} value of $10 \times \sigma_{\text{PSF}}$ and a \texttt{step-size} of $7\%$. Galaxy isophotes at the one and two effective radii are superimposed in red and blue, respectively. All images share the same color bar in units of \,$\text{counts s}^{-1}$, a size of $100 \times 100 \, \text{arcsec}^2$, and orientation (north up, east to the left). Galaxies are labeled with their corresponding names in gray boxes.}
        \label{fig:residuals}%
\end{figure*}

\begin{figure*}[t!] 
    \centering
    \includegraphics[bb=16 11 126 125,width=0.9\hsize,clip]{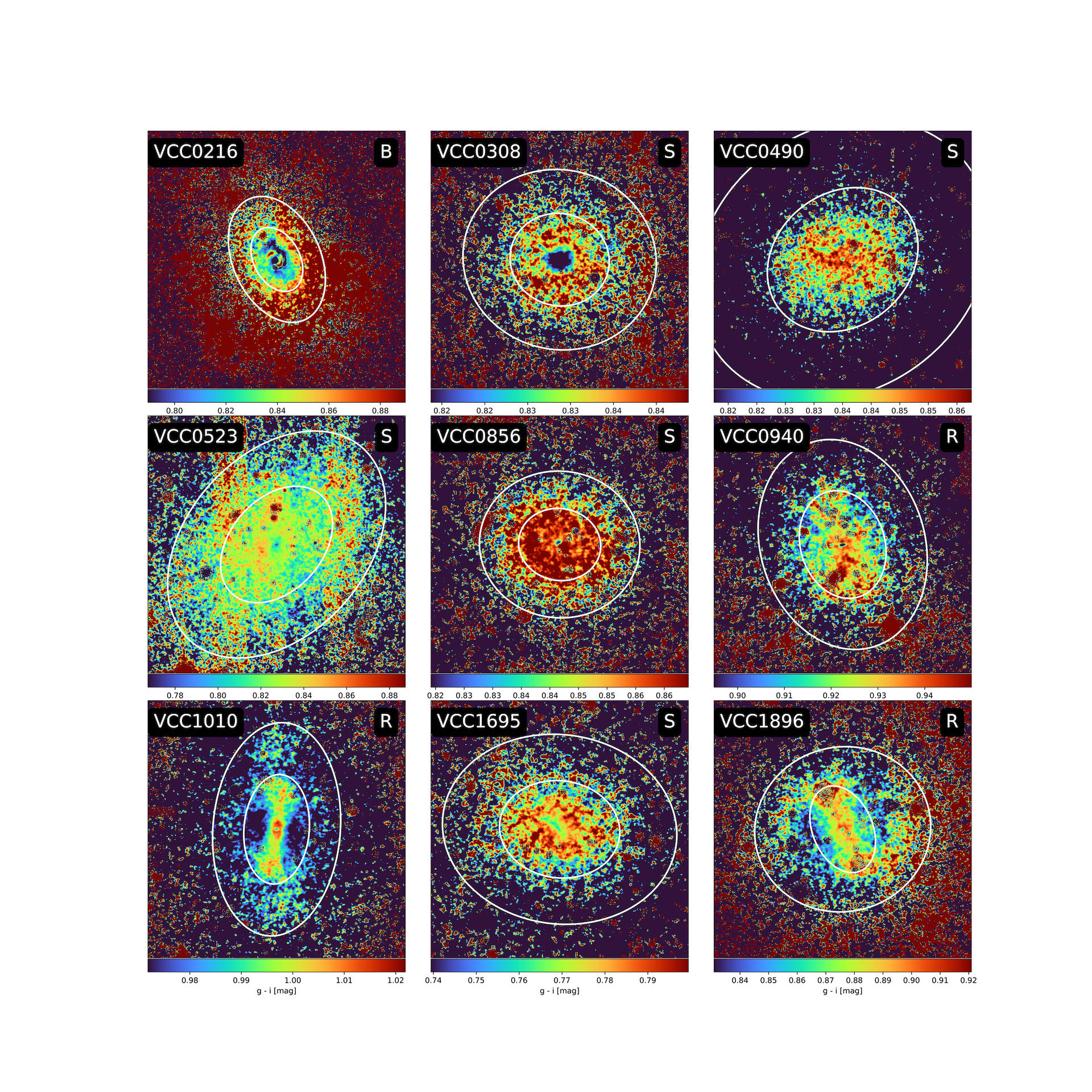}
    \caption{Color maps of the nine Virgo dwarf ETGs, each with its own color
        bar.  The maps show the \(g-i\) color in AB magnitude. after PSF matching, over a common \(100\times100~\text{arcsec}^2\) field with North up and East to the left.  Ellipses at one and two effective radii are overplotted in white. Letters in the top right of each panel denote spiral-like features that are redder than the diffuse component (R), similar in color to it (S), or bluer than it (B). Labels identify each galaxy; masked foreground objects appear as speckled regions.}
    \label{fig:color_map}
\end{figure*}

To quantify the luminosity of the spiral-like feature we measured the residual light fraction (RLF; \citealt{2021AJ....161..268M}), defined as the ratio of the residual image flux to the total galaxy flux inside the elliptical apertures of \(1\,R_{\mathrm{e}}\) and \(2\,R_{\mathrm{e}}\). Since we are working with 49 total combinations of smoothing and step-size, we end up with an equal number of residual images for each galaxy and exit criteria, thus we computed the RLF by taking the median of RLF value of the parameter distribution and derive the uncertainty calculating the $16^{\text{th}}$ and $84^{\text{th}}$ percentile of the distribution. Because the original JOI iterates longer than the new JOI, its residual images carry more flux and therefore yield larger RLF values (cf.\ Fig.~3 in \citealt{2021AJ....161..268M}). Considering that the new JOI is less susceptible to background compared to the original JOI, its excess light image is expected to be less susceptible to wrong arm-label pixels. Therefore, all subsequent measurements are calculated using the new JOI images and masks. Our resulting light fractions compiled in Fig.~\ref{fig:residual_fraction} span \(1.2\%\!-\!6.0\%\) (\(i\) band) and \(1.4\%\!-\!3.7\%\) (\(g\) band) within the \(1\,R_{\mathrm{e}}\) isophote, versus \(1.5\%\!-\!5.5\%\) (\(i\)) and \(1.7\%\!-\!3.7\%\) (\(g\)) within the \(2\,R_{\mathrm{e}}\) one. The close agreement between the one and two effective radii measurements suggests a lower contribution to the overall galaxy light by the arms beyond \(1\,R_{\mathrm{e}}\), as well as an equally reliable region to represent the spiral-like feature light fraction. 

Our RLF values are systematically lower than those of \citet{2021AJ....161..268M} as expected from the earlier termination of the revised jump-out criterion (Sect.~\ref{sec:new_joi}), which allows less flux to accumulate in the residual image. The offset is systematic rather than negligible; however, the relative ordering of the sample is preserved, and the galaxies with the largest and smallest light fractions are the same in both studies. We therefore regard the two sets of measurements as consistent in a relative sense, while noting that the absolute RLF is criterion dependent and should not be compared across implementations without accounting for this offset.

\begin{figure*}
        \centering
        \includegraphics[width=\hsize]{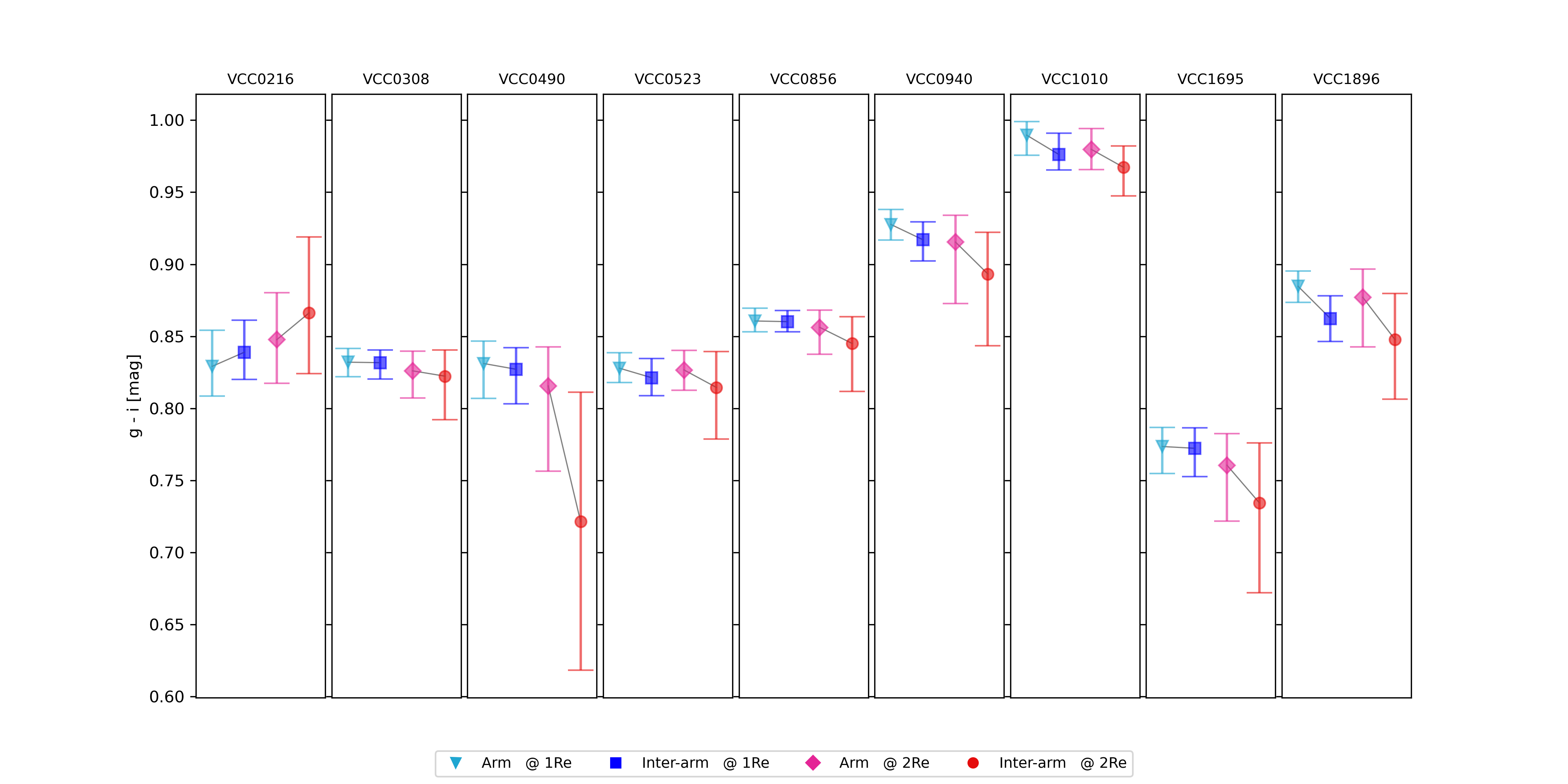}
        \caption{Color difference \(g-i\) of the interarm and arm pixels on the color map image within \(1\,R_{\mathrm{e}}\) and \(2\,R_{\mathrm{e}}\) of each Virgo dwarf. The median values within \(1\,R_{\mathrm{e}}\) of the arm and interarm regions are represented by a triangle and a square, respectively. The median values within \(2\,R_{\mathrm{e}}\) of the  arm and interarm   are represented by a diamond  and a circle respectively. A gray line connects the measurements of the arm and interarm regions of a galaxy at the same effective radius. The error bars are given by the $16^{\text{th}}$ and $84^{\text{th}}$ percentiles of the distribution.}
        \label{fig:color_map_measures}%
\end{figure*}

\section{Results}\label{sec:Results}

To investigate the stellar population in both spiral-like and diffuse structures, we examine the presence of color differences between the arm (spiral-like features) regions and the surrounding interarm (diffuse) component.

Fig.~\ref{fig:color_map} presents the \(g-i\) color maps for our sample. These maps were constructed by subtracting the PSF-matched \(i\)-band image from the corresponding \(g\)-band image, both in units of magnitude. The color scale is adjusted individually for each galaxy to enhance the visibility of potential structures. In several cases, the central color structure is almost entirely enclosed within the \(1\,R_{\mathrm{e}}\) ellipse (e.g., VCC\,0216, 0490, 0940, and 1695), consistent with the spatial extent of the substructures seen in the residual images (Fig.~\ref{fig:residuals}). An exception is VCC\,0523, whose color-defined structure extends out to approximately \(2\,R_{\mathrm{e}}\). For galaxies in which the substructures are not fully enclosed within \(1\,R_{\mathrm{e}}\) in the residual images, they are nevertheless contained within the \(2\,R_{\mathrm{e}}\) ellipse. Overall, most galaxies display a globally red color, with some exhibiting a bluer central region (e.g., VCC\,0308). Although the color maps are dominated by smooth galaxy light and do not always show sharply defined spiral-like features, comparison with the residual images reveals color variations spatially associated with spiral-like features in several ETGs. Clear examples include VCC\,0216, VCC\,1010, and VCC\,1896, where a small blue spiral pattern and two red bar-like structures are visible in the color maps. A less distinct case is VCC\,0940: while the spiral arms identified in the residual image appear blurred in the color map, their color is nevertheless visually redder than that of the interarm regions. For the remaining galaxies, no obvious color structure associated with the residual features is detected, including VCC\,0856, which shows prominent spiral arms in the residual image but lacks a corresponding color contrast.

Based on this visual inspection, we classified the galaxies into three categories: systems with redder spiral-like features (labeled "Redder"), systems with bluer spiral-like features  (labeled "Bluer"), and systems with no measurable color difference between their components (labeled "Similar"). These classifications are indicated in Fig.~\ref{fig:color_map}  with the letters "R","B", and "S", respectively.

Six galaxies (VCC\,0216, 0523, 0856, 0940, 1010, and 1896) show an unresolved blue feature at their center, consistent with the "nucleated" class of \citet{1985AJ.....90.1681B}. Although VCC\,1695 is listed as nonnucleated and was classified as similar in our visual inspection, it displays a central stripe suggestive of a weak nucleus \citep{Grant2005,2006ApJS..165...57C}, a nuclear bar, or material accreted onto the arms by the center. 

To quantify the color differences suggested by the maps in Fig.~\ref{fig:color_map}, we used the AMM and IAMM from the residual images to locate arm and interarm pixels on the color maps. Because the residual images are filter-based, the resulting masks can show small disagreements between filters; we therefore restrict the measurement to pixels common to both bands. Pixels flagged in the BPM (including the masked central region) are excluded from both the AMM and IAMM.

Figure~\ref{fig:color_map_measures} shows median \(g-i\)  color differences of nine galaxies in the arm  within \(1\,R_{\mathrm{e}}\) (triangle) and \(2\,R_{\mathrm{e}}\) (diamond), and  the interarm  \(1\,R_{\mathrm{e}}\) (square) and \(2\,R_{\mathrm{e}}\) (circle). The dominant trends are consistent with the visual classifications in Fig.~\ref{fig:color_map}. In particular, VCC\,0216 shows systematically bluer arms than interarm regions, while several galaxies show the opposite behavior, with arms redder than their interarm component (e.g., VCC\,0940, 1010, and 1896). Two systems (VCC\,0308 and 0856) show little or no arm and interarm offset, also in agreement with the previous classification.

At the same time, the arm and interarm color offsets are small compared to the width of the underlying pixel color distributions. The arm colors measured within $1\,R_{\mathrm{e}}$ and $2\,R_{\mathrm{e}}$ differ only weakly (typically by a few hundredths of a magnitude), consistent with most spiral-like features being  inside  the $1\,R_{\mathrm{e}}$ mark (Fig.~\ref{fig:residuals}). The interarm component shows a somewhat larger change between $1\,R_{\mathrm{e}}$ and $2\,R_{\mathrm{e}}$, with the outer regions generally becoming bluer, as expected if the diffuse light dominates at large radii but also from a stellar population gradient within the galaxy. Overall, the strong overlap between the arm and interarm distributions implies that the median differences alone do not provide compelling evidence for a statistically significant color separation. For completeness, in Fig.~\ref{fig:color_map_distribution} we present the full pixel color distributions of the arm and interarm regions, visually confirming this extensive overlap and illustrating the detailed shape of the data spread.

\begin{table*}
\caption{Galaxy arm color classification.}
\label{tab:significance_results}
\centering

\begin{tabular}{lccccc}
\hline
\noalign{\smallskip}

\text{VCC}  & Visual  & \multicolumn{2}{c}{P(arm>inter)} & S/N & Classification \\
&  & color map & residual & DSC & \\

(1)             &        (2)            &          (3)           & (4)  &   (5)  &   (6)  \\
\noalign{\smallskip}
\hline\hline
\noalign{\smallskip}

0216 & Bluer   &  $0.360^{+0.011}_{-0.004}$ & $0.401^{+0.049}_{-0.046}$ & 0.89  & Bluer     \\
0308 & Similar &  $0.558^{+0.005}_{-0.005}$ & $0.514^{+0.055}_{-0.082}$ & 7.55  & Similar   \\
0490 & Similar &  $0.799^{+0.005}_{-0.003}$ & $0.656^{+0.069}_{-0.065}$ & 0.00  & Similar   \\
0523 & Similar &  $0.660^{+0.009}_{-0.005}$ & $0.520^{+0.084}_{-0.064}$ & 6.73  & Similar   \\
0856 & Similar &  $0.655^{+0.007}_{-0.003}$ & $0.489^{+0.084}_{-0.082}$ & 0.35  & Similar   \\
0940 & Redder  &  $0.668^{+0.024}_{-0.016}$ & $0.627^{+0.053}_{-0.055}$ & 8.86  & Redder    \\
1010 & Redder  &  $0.734^{+0.006}_{-0.007}$ & $0.708^{+0.074}_{-0.043}$ & 17.08 & Redder    \\
1695 & Similar &  $0.663^{+0.003}_{-0.007}$ & $0.583^{+0.083}_{-0.067}$ & 10.97 & Redder    \\
1896 & Redder  &  $0.710^{+0.040}_{-0.012}$ & $0.826^{+0.058}_{-0.054}$ & 15.25 & Redder    \\

\noalign{\smallskip}
 
\hline
\end{tabular}
\tablefoot{Galaxy names are shown in column (1). Spiral-like feature color classification relative to the diffuse component based on visual inspection is reported in column (2); Column (3) and (4): Probability of picking an arm pixel redder than an interarm one on the color map and the residual-color map. The errors are represented by the $16^{\text{th}}$ and $84^{\text{th}}$ percentiles of the distribution. Column (5): S/N derived by comparing the DSC of the color-map arm mask with that of the residual arm mask, and the DSC noise distribution generated by comparing the residual AMM to an arm mask constructed from a color map in which pixels were shuffled independently within each isophote, if S/N \( > 1\) "Yes", otherwise "No". The final classifications are presented in column (6) following the criteria presented in Section \ref{subsec:color_clasification}.} 

\end{table*}

The width of these distributions also governs how the error bars in Fig.~\ref{fig:color_map_measures} are defined: we use the 16th and 84th percentiles to represent the spread of pixel values in each component (the central 68\% interval). These intervals quantify the intrinsic scatter of the color distribution, real radial gradients, incompletely subtracted diffuse light, and spatially correlated noise from PSF matching, rather than the uncertainty on the median itself. Because the measurements involve a very large number of correlated pixels, percentile-based spreads can remain broad even when coherent substructure is present, so these error bars should not be read as a direct measure of the significance of the difference between arm and interarm.

\begin{figure*}[h!]
        \centering
        \includegraphics[width=0.3\hsize]{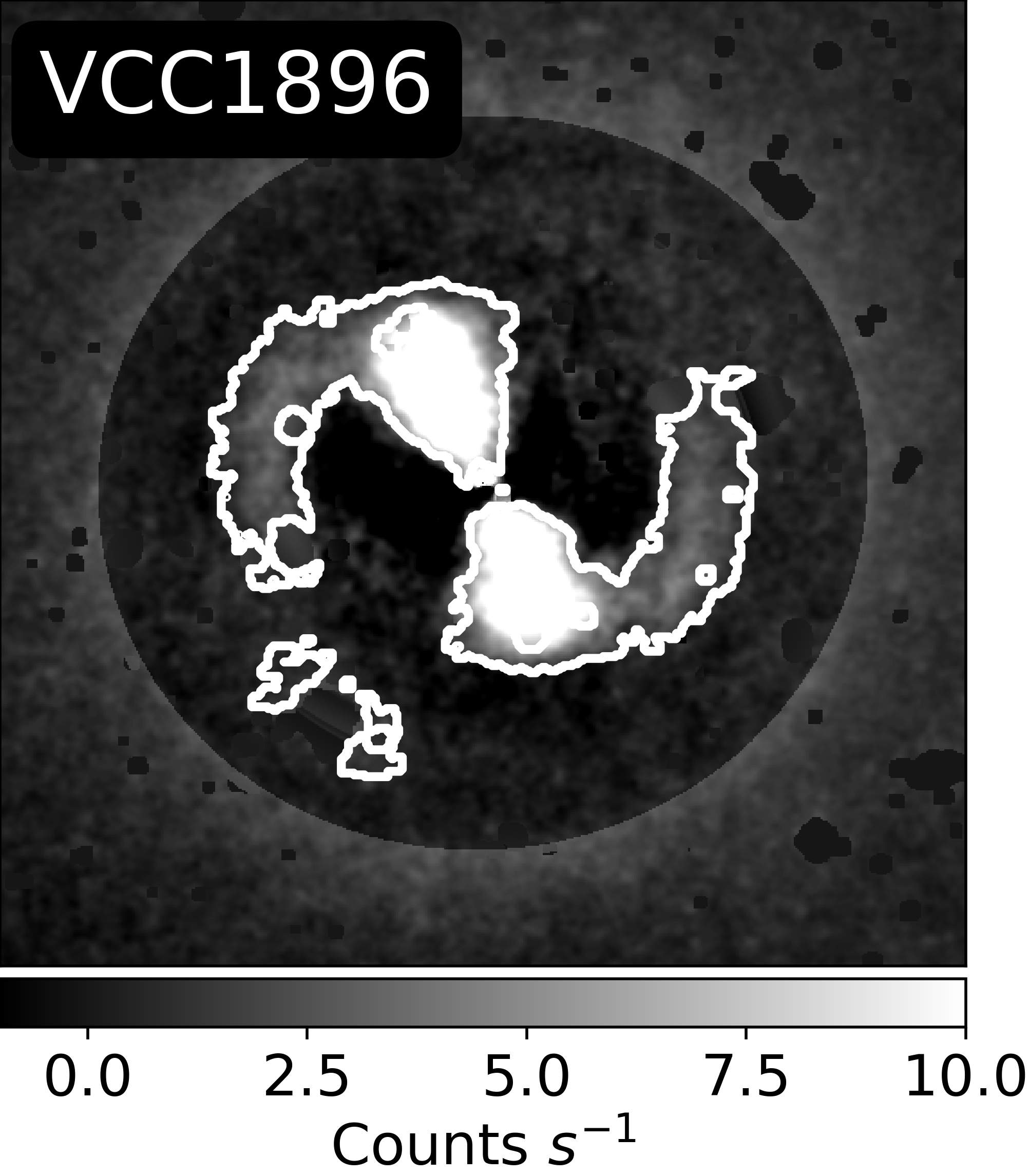}
        \includegraphics[width=0.3\hsize]{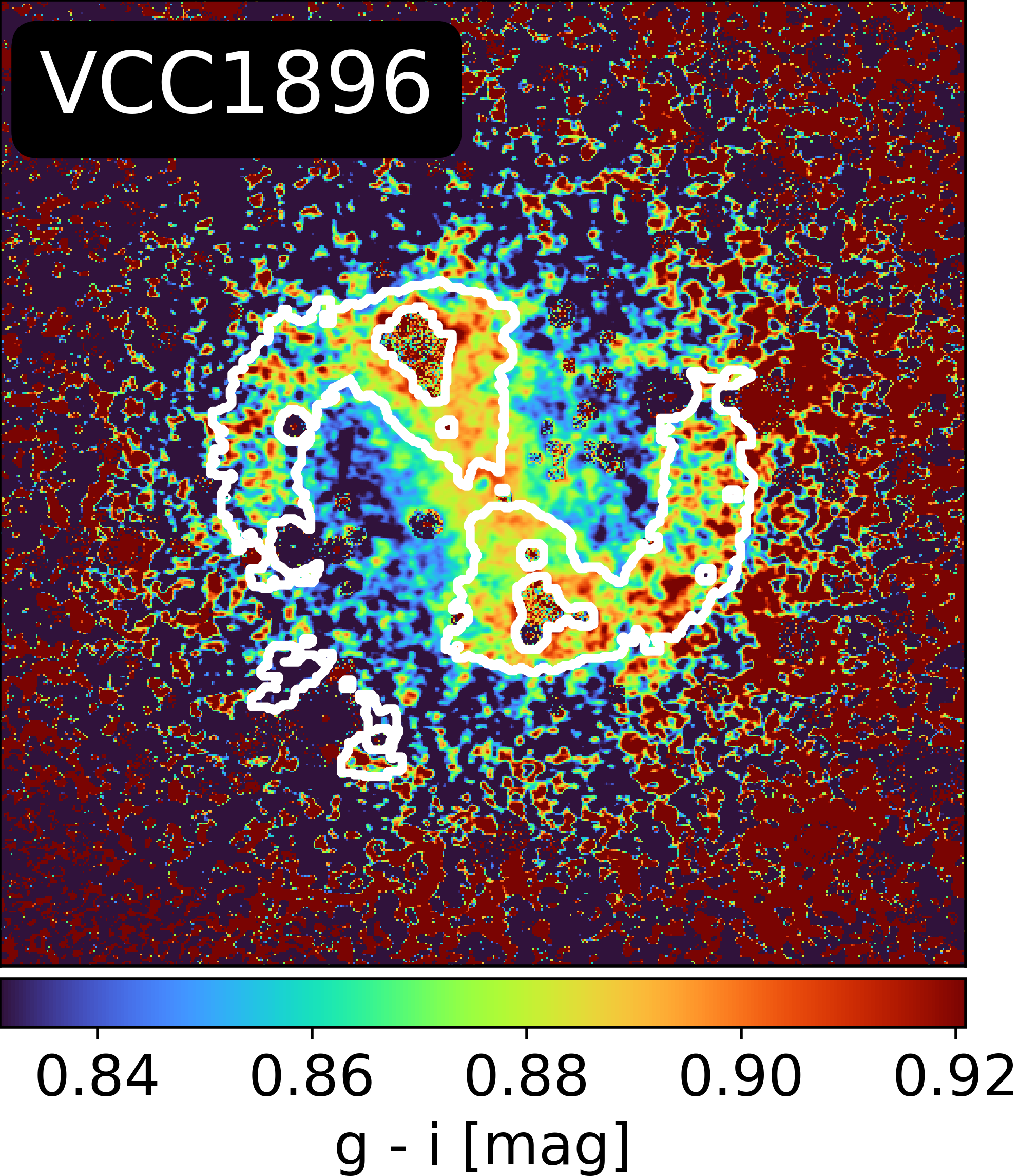}
        
        \caption{Residual image of VCC\,1896 with arm contours extracted from the residuals overplotted in white (left). Total \(g-i\) color map with the same contours overplotted (right), illustrating that the color map structure follows the residual spiral-arm morphology.}
        \label{fig:color_map_residual_resemblance}%
\end{figure*}

We tested whether a more sensitive measurement could be achieved by focusing on the excess light component traced by the residual image. Following the approach of \citet{2022AJ....164...18M}, we constructed residual and diffuse color-maps. The residual-color maps are derived from the residual images in each filter, while the diffuse-color maps are computed from images defined as the difference between the original and residual images in each filter. We then performed the measurements using the AMM and IAMM on both sets of maps. Unlike the total color maps, the residual-color maps measure the color of the excess light itself and are therefore not diluted by the diffuse host; the trade-off is a much broader per-pixel scatter, since the residual images are differences of two noisy images. The medians and $16^{\rm th}$--$84^{\rm th}$ percentile intervals shown in Fig.~\ref{fig:residual_color_map_measures} are consequently uninformative, for the same reason discussed above in connection with Fig.~\ref{fig:color_map_measures}, and we do not use them to assess significance. We instead characterize the residual-color maps with the probabilistic statistic $P(\mathrm{arm} > \mathrm{inter})$ described in Sect.~\ref{subsec:significance_estimation}, which is insensitive to the width of the underlying distributions; the resulting values are reported in column~4 of Table~\ref{tab:significance_results}.

Because the brightest spiral-like feature in our sample contributes only $\lesssim 6\%$ of the total galaxy light, any genuine color difference between the arm and the interarm regions, is inevitably diluted by the dominant diffuse component, and so the observed color differences are typically small ($\sim0.02 \text{mag}$), even when the true color difference is much larger. In the following section, we attempt to quantify the significance of the observed color differences between arm and interarm regions using  multiple approaches.

\subsection{Significance estimation of color differences}\label{subsec:significance_estimation}

The small differences between the median colors of the arm and interarm components shown in Fig.~\ref{fig:color_map_measures} are largely masked by the width of the corresponding distributions. However, this overlap alone does not imply that the two components are statistically identical. In some cases, the color map shows similar features to those in the residual image. For example, the red region of VCC\,1896 in Fig.~\ref{fig:color_map} shows a similar bar and spiral features as we see in its corresponding residual map (Fig.~\ref{fig:residuals}), highlighting this in Fig.~\ref{fig:color_map_residual_resemblance} where we overlap the contours from the residual image (left panel) on the color map (right panel). Similarly, the blue regions of VCC\,0216 show similar spiral arm features as in its corresponding residual map.

To assess this, we compared the color distributions of arm and interarm pixels within \(2\,R_{\mathrm{e}}\) to determine whether the observed difference in pixel color between the two regions is statistically significant. Pixel-based two-sample Kolmogorov--Smirnov \citep{kolmogorov1933sulla,smirnov1948table} and Mann--Whitney U tests \citep{mann1947test,wilcoxon1945individual} formally returned \(p\)-values $\leq10^{-5}$ for all galaxies (i.e., the difference in the distribution of interarm and arm pixels from the color map was statistically significant in all scenarios). However, these results are difficult to interpret because the pixel data are spatially correlated and comprise a very large number of samples. Under these conditions, standard nonparametric tests tend to reject the null hypothesis even for negligible differences and do not provide information on the direction of the effect. As such, they are not well suited for evaluating the significance of arm/interarm color differences. We therefore adopt two complementary approaches, one of which explicitly incorporates spatial information.

The first approach is a S/N estimator based on whether the structures observed in the color maps (Fig.~\ref{fig:color_map}) correspond to those seen in the residual images (Fig.~\ref{fig:residuals}). To account for the full spatial extent of the substructures, we restrict all measurements to within $2\,R_{\mathrm{e}}$. To construct a mask of coherent color structures, we placed a series of isocolor contours on the color map, with values ranging from the 5th to the 95th percentile of the color-pixel distribution in steps of $0.01\,\mathrm{mag}$. We then search for the isocolor contour that provides the closest match to the AMM (arm measuring mask derived from the residual images). The agreement is quantified using the Dice Similarity Coefficient \citep[DSC;][]{https://doi.org/10.2307/1932409}, defined as
\begin{equation}
\mathrm{DSC} = \frac{2\,|A \cap B|}{|A| + |B|},
\end{equation}
where $A$ is the set of pixels enclosed by a given isocolor contour and $B$ is the set of pixels belonging to the AMM. The coefficient is $1$ for a perfect match and $0$ for no overlap, and the isocolor contour with the highest DSC is selected as the closest color-map match to the residual features.

To determine whether the observed DSC could arise by chance, we built a null model to quantify its significance. We randomized the pixels between the arm and interarm regions at a fixed radius (shuffling labels along fixed isophotes). This procedure preserves the global morphology but effectively removes any true color difference between the arm and interarm regions. We then remeasured the DSC on the randomized color map. Repeating this randomization procedure $10^{4}$ times yields a null distribution of DSC values, from which we obtain the mean $\langle \mathrm{DSC}\rangle_{\mathrm{rand}}$ and standard deviation $\sigma_{\mathrm{DSC}}$ of the DSC for images in which the arm--interarm color differences have been erased.
The signal is simply the DSC measured in the galaxy, $\mathrm{DSC}_{\mathrm{sci}}$. To quantify its significance, we measure where this value lies with respect to the mean and standard deviation of the randomized-pixel null distribution, defining the S/N as
\begin{equation}
S/N = \frac{\mathrm{DSC}_{\mathrm{sci}} - \langle \mathrm{DSC}\rangle_{\mathrm{rand}}}{\sigma_{\mathrm{DSC-rand}}} .
\end{equation}
This measures how far the observed DSC lies above the randomized (noise) distribution, in units of its dispersion. We use this $S/N$ to evaluate the significance of the observed arm and interarm color difference.

For the second approach to measure the significance of the color difference, we tested the arm and interarm difference using an alternative probabilistic color statistic, $P(\mathrm{arm} > \mathrm{inter})$, defined as the probability that a randomly selected arm pixel is redder than a randomly selected interarm pixel; equivalently, it measures how often arm pixels have higher color values than interarm pixels when comparing random pairs. A value of \(P=0.5\) indicates there is no color difference between the two components, whereas \(P>0.5\) indicates a redder arm component and \(P<0.5\) indicates a bluer one. For example, \(P \approx 0.8\) corresponds to a strongly redder arm component, while \(P \approx 0.2\) corresponds to a strongly bluer one. We measured $P(\mathrm{arm} > \mathrm{inter})$ in both the color map and the residual-color map. The resulting S/N values and color probabilities are reported in Table~\ref{tab:significance_results}.

We now summarize the results of the significance analysis presented in Table~\ref{tab:significance_results} and discuss their implications for the arm and interarm color classification. Galaxies visually classified as having redder arm structures (VCC\,0940, 1010, and 1896) exhibit strong correspondence between the shapes seen in the color maps and those in the residual images, with $\mathrm{S/N}>8$. Systems with weak or no apparent color structure (e.g., VCC\,0490 and 0856) show correspondingly low values ($\mathrm{S/N}\leq1$). The only visually blue system, VCC\,0216, also yields a weak structural signal, consistent with the low-contrast spiral pattern in the color map.

The residual-color probabilities behave as expected of a measurement free of dilution by the host. They retain their offset from $P = 0.5$ in the systems classified as redder (VCC\,0940, 1010, 1695, and 1896) and in the single bluer system (VCC\,0216), while falling to within ${\sim}0.02$ of the null value in the three systems with no genuine contrast (VCC\,0308, 0523, and 0856). For VCC\,1896, the galaxy with the most prominent features, the residual-color probability exceeds that measured on the total color map ($0.826$ versus $0.710$), consistent with the latter being suppressed by the dominant diffuse component.

\subsection{Final color difference classification}\label{subsec:color_clasification}

To consolidate the results into a single color classification (redder, bluer, or similar), we combined the outcomes of three independent diagnostics: (i) visual inspection of the color map, (ii) the color probability statistic measured from the color map, and (iii) the same probability measured from the residual-color map. Based on these diagnostics, we defined the following classification criteria.

If all three methods agree, the galaxy is assigned the corresponding classification. If only two of the three diagnostics agree, we then evaluate the S/N of the color difference. When the S/N is significant (defined as \(\mathrm{S/N} > 1\)), the galaxy is classified according to the most frequent outcome among the three diagnostics. In all remaining cases, where the diagnostics disagree and the S/N is not significant, we conservatively classified the galaxy as "similar", indicating no robust color distinction between the spiral-like feature and diffuse components. The final classifications are listed in Column~6 of Table~\ref{tab:significance_results}.

Overall, most galaxies retain their original visual classification. One notable exception is VCC\,1695, which shifts from similar to redder under the combined criteria. This reclassification yields equal numbers of galaxies with spiral-like features redder than their diffuse component and galaxies showing no significant color difference between the two components.

\section{Discussion}\label{sec:Discussion}

We studied the nine Virgo dwarf ETGs targeted by \citet{2021AJ....161..268M}, using deep \(g-i\) color imaging to try to measure the optical colors of their hidden spiral-like features, in order to better constrain their origins. We evaluate the significance of the observed color differences using multiple statistical approaches. We find that four dwarfs host spiral-like features redder than their diffuse component, one shows a bluer feature, and four present no significant color difference between the two components. In this section we examined how the different proposed origin scenarios could each produce the color differences observed in our dwarfs.

\subsection{What are the implications of color differences between the spiral-like and diffuse components?}\label{subsec: discuss. 1}

Following the structural diversity observed in our sample spiral-like features, we find an analogous diversity in color behavior for these structures. Not all galaxies display the same trend: four systems exhibit spiral-like features redder than their diffuse surroundings, one galaxy shows a bluer spiral-like features, and the remaining four present no significant color difference between the two components. This diversity suggests that multiple evolutionary pathways can produce similar present-day morphologies within the Virgo cluster.

The color contrast between the spiral-like and diffuse components can serve as a tracer of their relative stellar population ages and star-formation histories. If a galaxy has fallen into the cluster only recently and experienced rapid quenching due to environmental mechanisms such as ram-pressure stripping or harassment, we would expect the spiral-like feature that originally hosted most of the star formation to appear bluer than the diffuse component (i.e., VCC\,0216, and Fig.~\ref{fig:color_map}). A similar outcome could occur if gas and/or younger stellar material were deposited into the spiral-like feature region through minor mergers or tidal interactions, leading again to a visibly younger spiral-like feature relative to its surroundings.

On the other hand, if star formation ceased several gigayears ago, both components may have aged sufficiently that any initial differences have faded. In this case, the spiral-like feature and diffuse light would exhibit similar colors, consistent with coeval quenching or a long period of passive evolution during which their stellar populations evolved in parallel. Galaxies in such a state may retain spiral-like features, but no longer display strong color gradients capable of revealing them photometrically. 

However, \citet{2021ApJ...912..149S} find that similar spiral-like features can be tidally triggered by plunging pericenter passages ($r_{\mathrm{peri}} < 0.25\,r_{\mathrm{vir}}$) if a galaxy contains a cold, thin disk embedded within a more massive and dynamically hot disk. In this scenario, tidal perturbations excite spiral-like features in the cold component while leaving the dominant hot component smooth and featureless. Unfortunately, photometry alone cannot determine the relative rotation velocities of the cold and hot components, and such transition dwarfs may not necessarily exhibit strong color gradients.

Alternatively, the stellar populations may not differ primarily in age but instead in metallicity. If the stars in the spiral-like feature are more metal-rich than those in the diffuse component, the spiral-like feature may appear redder despite having formed at a similar epoch. In this case, a tidally triggered spiral feature may appear redder despite having similar ages to the surrounding stellar population. Such metallicity-driven color differences have been observed in some early--type dwarfs. For example, \citet{2017A&A...606A.135U} and \citet{2022MNRAS.515.4622B} find that certain Virgo dETGs host young but metal-rich stellar populations in their central regions. If an embedded cold disk is rotationally supported and metal-rich, this scenario could naturally explain the red bar systems observed in VCC\,0940, VCC\,1010, and VCC\,1896.

\subsection{How can we further constrain the origin of embedded spiral-like features?}\label{subsec: discuss. 2}

A more complete understanding of the origin of embedded spiral-like features in dwarf ETGs requires both deeper observational constraints and a broader statistical context. In particular, spatially resolved spectroscopy is needed to measure stellar ages, metallicities, and kinematics of the substructures independently from the diffuse host. At the same time, a substantially larger and less biased sample of dwarf ETGs is required in order to robustly assess how disk properties depend on galaxy mass, cluster-centric position, and substructure morphology, and to place the Virgo systems within the wider diversity of dwarf galaxy evolutionary pathways across environments.

It is important to note that our sample is limited both in size (nine dwarf ETGs) and in selection. The galaxies were originally chosen for the presence of spiral-like features identified in SDSS imaging by \citet{2006AJ....132..497L}, which biases the sample toward systems with relatively prominent spiral-like features. This precludes a systematic exploration of the full range of embedded features and may limit the diversity of stellar population signatures detected in our analysis. In contrast, \citet{2022AJ....164...18M} studied a larger sample of 23 dwarf ETGs in the Fornax cluster selected using a photometric–morphological approach that naturally included a broader variety of systems. Their work distinguished between "clumpy" and "disky" subclasses and demonstrate that these groups follow different evolutionary paths. Such a separation is not feasible with our limited sample, which may explain why we identify only one galaxy with a clearly bluer spiral-like feature, whereas a larger or more representative sample could reveal a larger number of bluer spiral-like features and provide the statistical weight needed to characterize the colors of the spiral-like features robustly.

Several galaxies in our sample have been observed spectroscopically in previous studies (e.g., \citealp{2016ApJ...818..179L,2017MNRAS.470..815S,2025ApJ...979...85L}). However, these data generally rely on fixed apertures or integrated spectra that are dominated by the diffuse stellar component. As a result, they provide limited information on the embedded spiral-like features themselves, which contribute only a small fraction of the total galaxy light. Consequently, stellar population properties derived from such spectra cannot be directly associated with the spiral-like features identified in imaging.

An important exception is the work by \citet{2025ApJ...979...85L}, who separated the central spectra of dwarf ETGs into contributions from the nuclear star cluster (NSC) and the surrounding host region. For two of our similar galaxies, the NSC is younger and more metal-rich than the host, while another similar system shows an older and more metal-poor NSC compared to its diffuse component. One of our "red" galaxies, VCC\,1896, was observed with MUSE and integral field spectra were analyzed by \citet{2023MNRAS.525.4329B},  who derived the 2D spatial distribution of the galaxy stellar age and metallicity. These maps show that the galaxy is rather uniformly old through its main body, while its central bar is more metal-rich than the surroundings. The studies of \citet{2025ApJ...979...85L} and \citet{2023MNRAS.525.4329B} highlight the significant diversity in the star formation histories of dwarf ETGs and underscore the limitations of using central or integrated spectra to infer the properties of faint, spatially extended features.

The \(g-i\) colors measured here reveal that, in five of the nine galaxies, the stellar populations of the spiral-like features differ from those of the main galaxy body, which already places some constraints on their origin. It is difficult, however, to distinguish between a difference in age and a difference in metallicity using a broadband color alone. In the long term, spatially resolved spectroscopy of the arm and interarm regions is the natural route to resolving this ambiguity. Spectroscopy alone, however, does not remove the underlying difficulty: the features contribute only a few percent of the total light, so a spectrum extracted at their position is likewise completely dominated by the diffuse host, with the added requirement that the feature must first be located, which is not reliably achievable by visual inspection. In two of our nine galaxies the decomposition does not reproduce what the eye suggests: VCC\,0856 shows prominent arms in the residual image (Fig.~\ref{fig:residuals}) yet no measurable color contrast (S/N $= 0.35$; Table~\ref{tab:significance_results}), while VCC\,1695, classified as similar by eye, is classified as redder once the diagnostics are combined (S/N $= 10.97$; Table~\ref{tab:significance_results}).

The residual method can help in this respect. It identifies the exact regions of the dwarf disk in which the features are located and quantifies their contribution to the total light of the galaxy in that region; the RLF measured in Sect.~\ref{sec:Residual_Method} range from approximately 1.5\% to 5.5\% within $2\,R_{\mathrm{e}}$, with VCC\,1896 the most prominent case and VCC\,0856 among the faintest (Fig.~\ref{fig:residual_fraction}). Spectroscopic follow-up can then be designed for the maximum chance of success and the most efficient use of telescope time, whether by placing slits in the optimal location or by selecting which IFU spaxels to stack. A methodology for separating the spectral flux of the spiral-like features from that of the diffuse component can likewise be designed in advance, with prior knowledge of the expected relative contributions of the two components and of whether emission lines might or might not be present. We therefore consider deep multiwavelength imaging a highly useful and complementary preliminary step for spectroscopic studies of these features.

A further consideration is that photometric methods of this kind scale more readily than spectroscopy alone. The Legacy Survey of Space and Time (LSST) at the Vera C. Rubin Observatory \citep{2019ApJ...873..111I} began in 2026 and, together with Euclid \citep{2011arXiv1110.3193L} and The Nancy Grace Roman Space Telescope \citep{2019arXiv190205569A} , will deliver deep wide-field imaging of nearby clusters in samples orders of magnitude larger than any assembled to date. The spectroscopy of these surveys is optimized for redshift measurement rather than for resolved stellar populations in low-surface-brightness structures. If embedded substructures are to be characterized in these datasets at all, it will initially have to be done photometrically. This study therefore presents the development and validation of an approach, tested on a well-characterized sample with deep, high-quality imaging, that could potentially be automated in the near future for application to much larger samples.

\section{Summary and conclusions}\label{sec:Summary}

This work is a pilot study establishing a multiband extension of the residual-method technique of \citet{2021AJ....161..268M} on a well-characterized sample of nine Virgo dETGs with previously identified embedded spiral-like features, but without color information. Using deep NGVS \(g\)-and \(i\)-band imaging we (i) Recover the same range of residual light fractions from those reported in \citet{2021AJ....161..268M}, with a systematic offset attributable to the revised stopping criterion. Finding \(1.5\%\!-\!5.5\%\) in \(i\) and \(1.7\%\!-\!3.7\%\) in \(g\) within $2\,R_{\mathrm{e}}$; (ii) extend the analysis to \(g-i\) colors for both the spiral-like features and their diffuse hosts; and (iii) demonstrate the readiness of the technique for the much larger NGVS sample.

Using \(g-i\) color maps and residual-color maps with significance estimators that incorporate spatial coherence, we classify the sample into four galaxies with redder spiral-like features, four with no robust color distinction, and one with a bluer feature. This diversity may reflect multiple evolutionary pathways: the single bluer feature is consistent with recent quenching during cluster infall, as is frequently anticipated for these structures; systems where no color difference is detected host older, coevally quenched features whose initial contrasts have faded; redder features may arise if an embedded, dynamically cold component is more metal-rich and/or older than the surrounding diffuse stars. Notably, only one of the nine galaxies hosts a bluer spiral-like feature, fewer than would be expected if these structures were predominantly the fossil arms of recently quenched infalling dwarf spirals. Whether this reflects the intrinsic population or the limited size and morphology-based selection of our sample can only be established with the larger, less biased sample discussed below.

Breaking the age--metallicity degeneracy with deep optical imaging alone remains difficult, and spatially resolved spectroscopy is the natural route to resolving it. Neither approach, however, stands on its own. Because the spiral-like features contribute only a few percent of the total light, any spectroscopic measurement is subject to the same dilution that limits our colors, with the additional requirement that the features first be located, which is precisely what the residual method provides. The two are therefore complementary rather than alternative.

The nine galaxies are a small, morphology-selected subset of a much larger parent population, the Virgo Cluster Catalog \citep{1985AJ.....90.1681B} contains approximately $2000$ probable members, and visual classification of NGVS imaging identifies over $100$ dwarfs with secure or possible spiral-like substructures \citep{2025AJ....170..363K}. Our pilot results motivate a forthcoming systematic analysis of this larger sample that would enable correlations not accessible here: substructure morphology versus cluster-centric position, galaxy stellar mass versus substructure prominence, and color versus substructure shape. 

\begin{acknowledgements}

RS acknowledges financial support from FONDECYT Regular 2023 project No.~1230441, and also gratefully acknowledges financial support from ANID-MILENIO NCN2024--112. This research was supported by the Korea Astronomy and Space Science Institute under the R\&D program (Project No.~2026--1--831--03) supervised by the Korea AeroSpace Administration.

Based on observations obtained at the Canada-France-Hawai'i Telescope (CFHT), which is operated by the National Research Council of Canada, the Institut National des Sciences de l'Univers of the Centre National de la Recherche Scientifique of France, and the University of Hawai'i. CFHT is located on Maunakea on Hawai'i Island, a mountain of considerable cultural, natural, and ecological significance. Maunakea is a sacred site to Native Hawaiians, also known as Kānaka 'Ōiwi. We would like to thank the CFHT Telescope Operations and Software Groups for their contributions and diligence in maintaining observatory operations; the CFHT Astronomy Group for their observation coordination and data acquisition efforts; and the CFHT Finance \& Administration Group for their contributions to the management and administration of the observatory. Based on observations obtained with MegaPrime/MegaCam, a joint project of CFHT and CEA/DAPNIA.

This research used the Canadian Advanced Network for Astronomy Research (CANFAR), operated in partnership by the Canadian Astronomy Data Center and the Digital Research Alliance of Canada with support from the National Research Council of Canada, the Canadian Space Agency, CANARIE, and the Canadian Foundation for Innovation.

\end{acknowledgements}

\bibliographystyle{aa}
\bibliography{reference_file}


\begin{appendix}
\onecolumn

\section{Construction of the refined arm measuring mask}\label{apx:new_mask_steps}

This appendix describes the construction of the refined Arm Measuring Mask (AMM) adopted by the new jump-out criterion of Sect.~\ref{sec:new_joi}; the interarm Measuring Mask (IAMM) is its complement within the valid galaxy footprint. The goal is to identify pixels associated with coherent embedded spiral-like substructures (spiral arms or bars) so that the remaining nonarm pixels, a mixture of incompletely subtracted diffuse-component light and sky-level fluctuations, can be used to monitor convergence.

\paragraph{Inputs and definitions.}
The procedure makes use of: (i) the iteration-0 (i.e., the first run of the residual method) residual image, (ii) the bad-pixel and foreground mask (BPM), (iii) a binary galaxy-footprint mask defined by the outermost valid isophote, and (iv) the full width at half maximum of the point spread function, $\mathrm{PSF}_{\mathrm{FWHM}}$, which characterizes the effective spatial resolution (seeing) of the image. Structures significantly smaller than the PSF are unresolved and are therefore treated as noise-dominated.

All operations are restricted to the galaxy footprint and exclude BPM-masked pixels, as well as the central region excluded from the analysis. We defined positive residual pixels as those with residual values $>0$, corresponding to excess light relative to the smooth host-galaxy model.

\paragraph{Step-by-step procedure.}
\begin{enumerate}
  \item Initial detection (\textsc{SExtractor})\,: We run \textsc{SExtractor} on the iteration-0 residual image to generate an initial \texttt{SEGMENTATION} map. A \texttt{WEIGHT\_IMAGE} is supplied such that only pixels within the galaxy footprint and outside the BPM are considered. A deliberately low detection threshold of $\mathrm{S/N}=1$ is adopted to retain faint arm fragments that may lie close to the noise level.%
  \footnote{The most relevant \textsc{SExtractor} parameters for this step are \texttt{DETECT\_THRESH}/\texttt{ANALYSIS\_THRESH}, \texttt{DETECT\_MINAREA}, and the deblending parameters \texttt{DEBLEND\_NTHRESH} and \texttt{DEBLEND\_MINCONT}.}

  \item Morphological opening\,: The segmentation map is converted into a binary structure mask. A morphological \texttt{BINARY OPENING} operation with an elliptical structuring element is applied to remove isolated noise pixels while preserving spatially coherent detections.

  \item Size filtering\,: Connected components with effective diameters smaller than $1.5 \times \mathrm{PSF}_{\mathrm{FWHM}}$ are removed. This suppresses unresolved features that are comparable to or smaller than the image resolution and are therefore likely dominated by noise.

  \item Positive-residual growth\,: To recover low-S/N arm pixels that may have been missed or eroded in previous steps, we expand the cleaned structure mask by reinserting all positive residual pixels located within a radius of 5 pixels from the current mask boundary. Restricting this recovery step to positive residuals avoids incorporating negative troughs that can arise from over-subtraction or background mismatches.

  \item Smooth open/close\,: A second \texttt{OPENING} followed by a \texttt{CLOSING} operation is applied to smooth mask boundaries, fill small holes, and remove residual speckles while maintaining the connectivity of arm structures.

  \item Consistency constraint\,: The refined structure mask is intersected with the original \textsc{SExtractor} segmentation map to ensure that all retained regions remain anchored to detections in the residual image.

  \item Final cleanup\,: The size-filtering criterion of step~3 is repeated to remove any small artifacts introduced by the recovery and smoothing operations.

  \item Lost-pixel compensation\,: A final footprint recovery step, analogous to step~4 but using a smaller radius of 3 pixels, is applied to compensate for minor pixel losses caused by morphological erosion, ensuring that contiguous arm structures remain fully covered.
\end{enumerate}

\paragraph{Final interarm mask.}
The refined interarm measuring mask is defined as
$$\mathrm{IAMM} = \mathrm{Footprint} \setminus \left(\mathrm{Structure\ Mask} \cup \mathrm{BPM}\right)$$
where the Structure Mask is the final product of steps~1--8 above. The corresponding refined arm measuring mask (AMM) is given by the structure mask itself, restricted to the same footprint and BPM exclusions.

\paragraph{Robustness checks.}
We verified that the resulting IAMM and AMM masks are stable against small variations in the key hyperparameters (e.g., recovery radius varied by $\pm2$ pixels and detection thresholds in the range $\sim0.8$--$1.2\sigma$). The inferred arm and interarm color trends remain unchanged within the quoted uncertainties.

\begin{figure*}[h] 
\centering

\begin{minipage}[t]{0.23\textwidth}
  \centering
  \includegraphics[width=\linewidth]{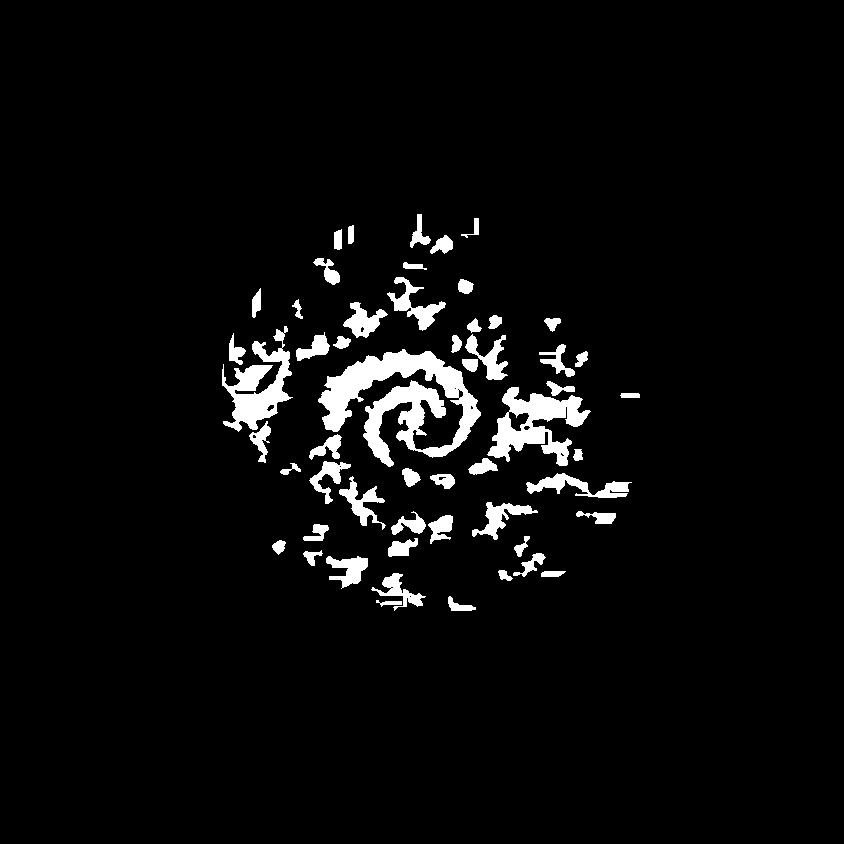}\\
  \small (a) Step 1: \textsc{SExtractor} segmentation.
\end{minipage}\hfill
\begin{minipage}[t]{0.23\textwidth}
  \centering
  \includegraphics[width=\linewidth]{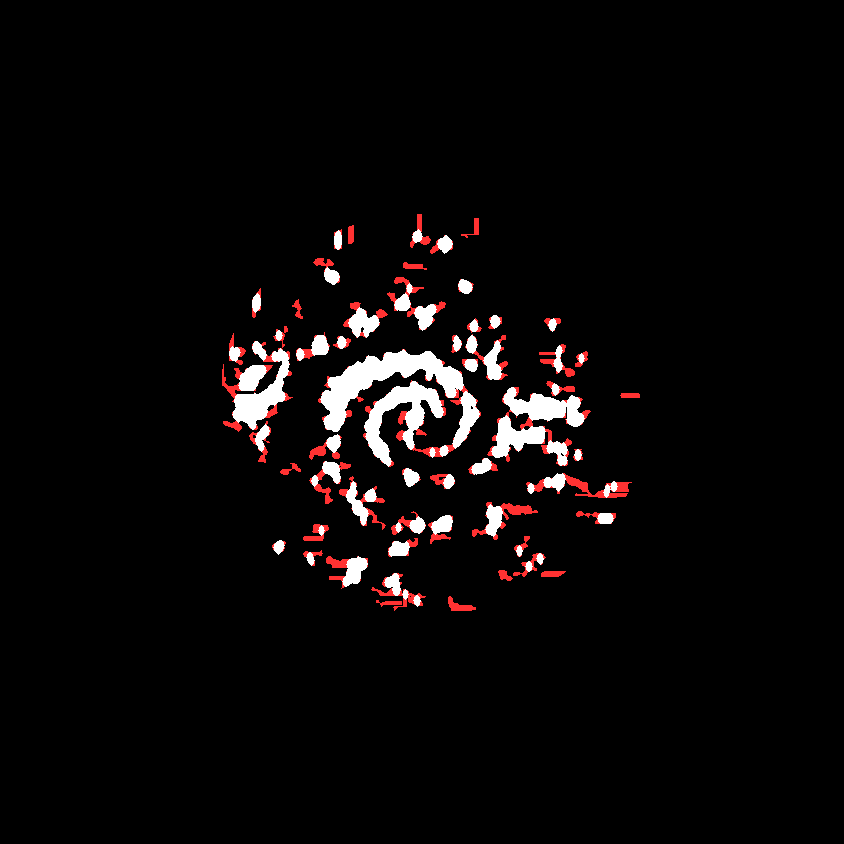}\\
  \small (b) Step 2: binary opening.
\end{minipage}\hfill
\begin{minipage}[t]{0.23\textwidth}
  \centering
  \includegraphics[width=\linewidth]{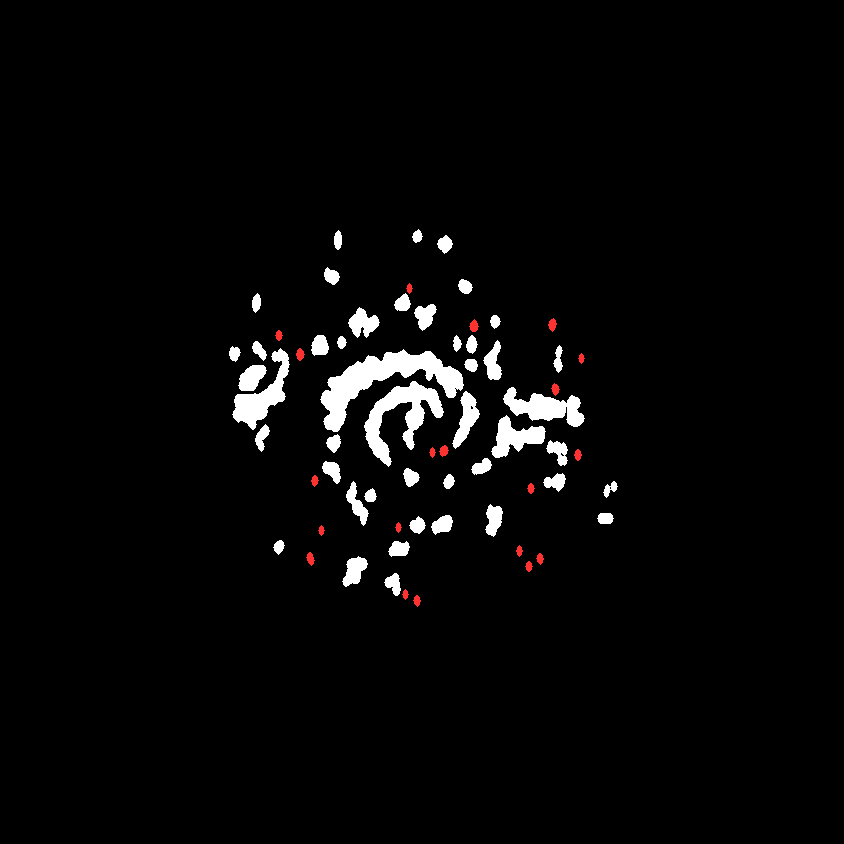}\\
  \small (c) Step 3: size filtering.
\end{minipage}\hfill
\begin{minipage}[t]{0.23\textwidth}
  \centering
  \includegraphics[width=\linewidth]{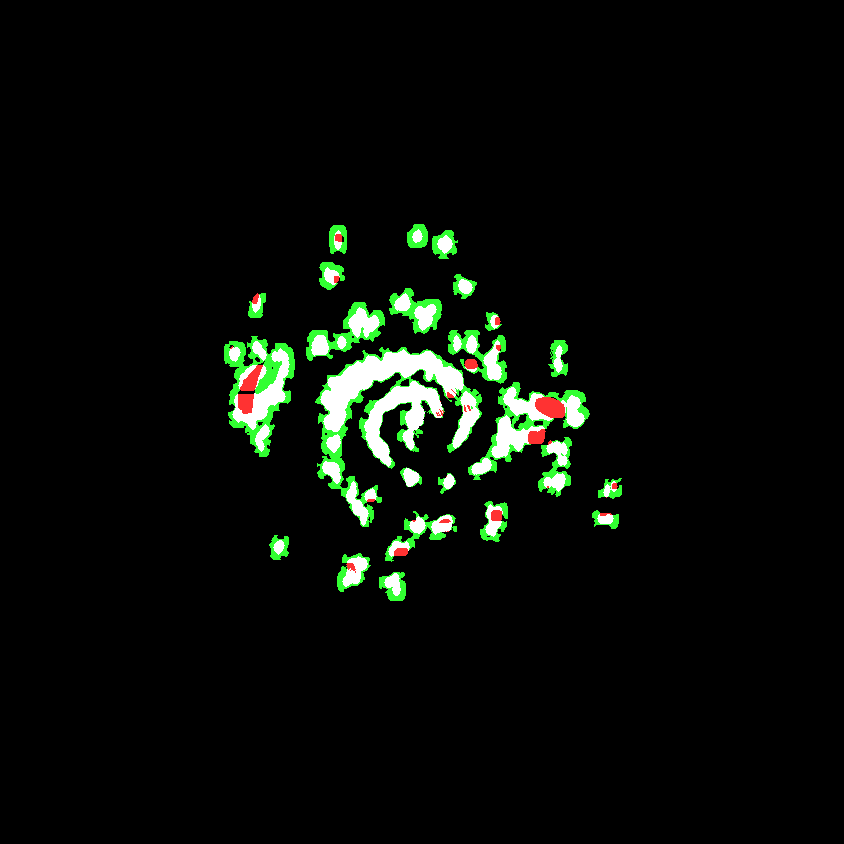}\\
  \small (d) Step 4: positive-residual growth.
\end{minipage}

\vspace{3mm}

\begin{minipage}[t]{0.23\textwidth}
  \centering
  \includegraphics[width=\linewidth]{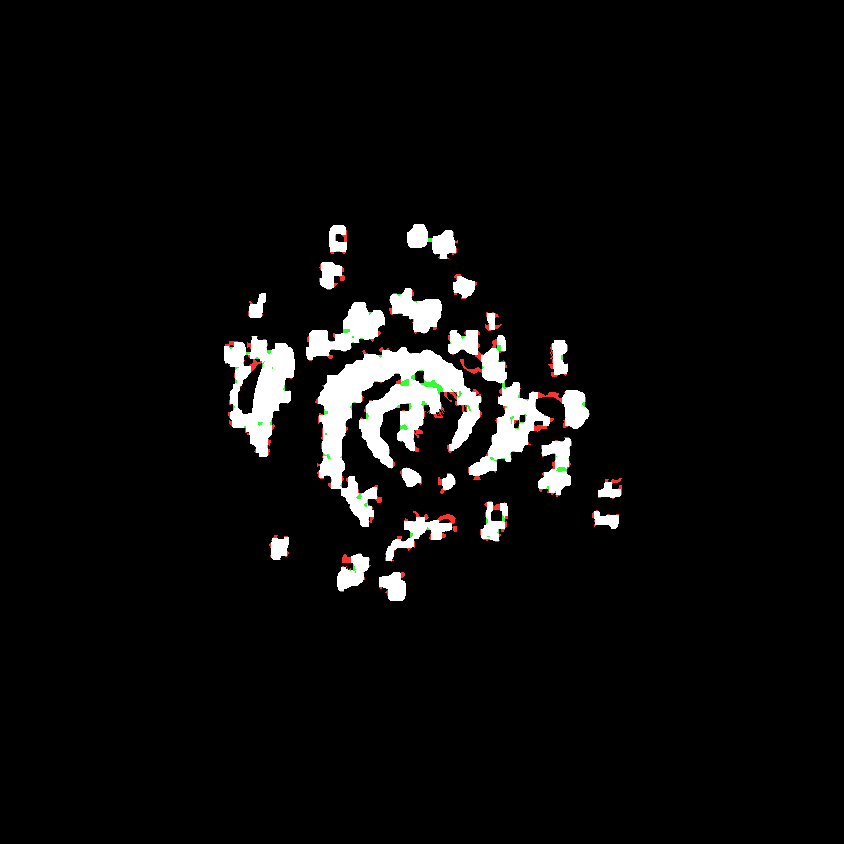}\\
  \small (e) Step 5: edge smoothing (opening+closing).
\end{minipage}\hfill
\begin{minipage}[t]{0.23\textwidth}
  \centering
  \includegraphics[width=\linewidth]{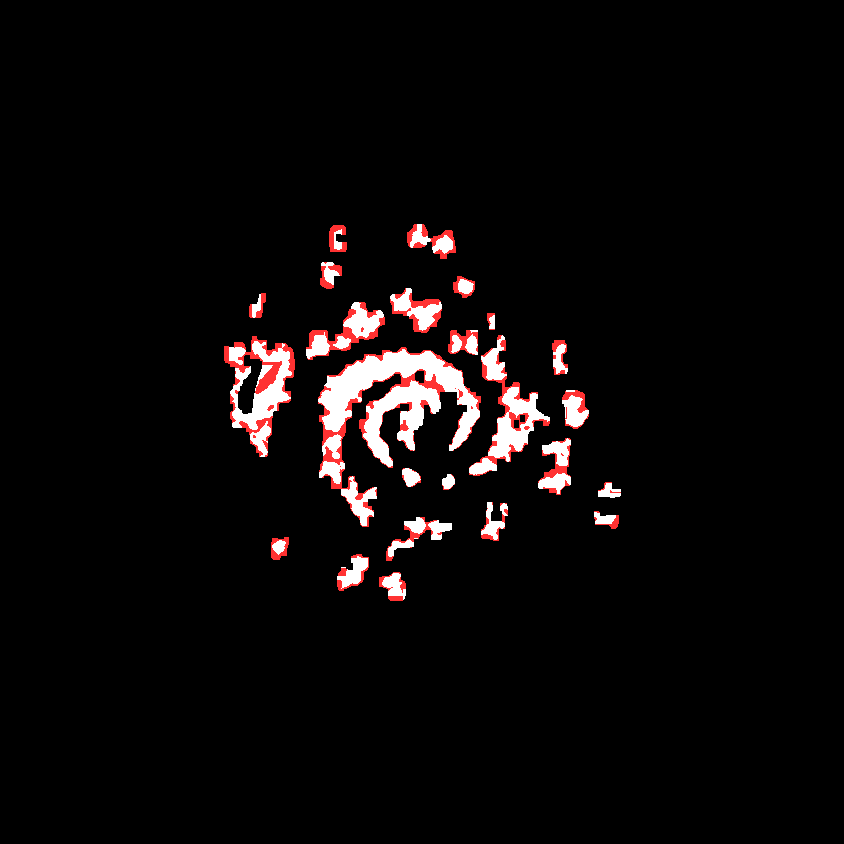}\\
  \small (f) Step 6: consistency constraint (intersection).
\end{minipage}\hfill
\begin{minipage}[t]{0.23\textwidth}
  \centering
  \includegraphics[width=\linewidth]{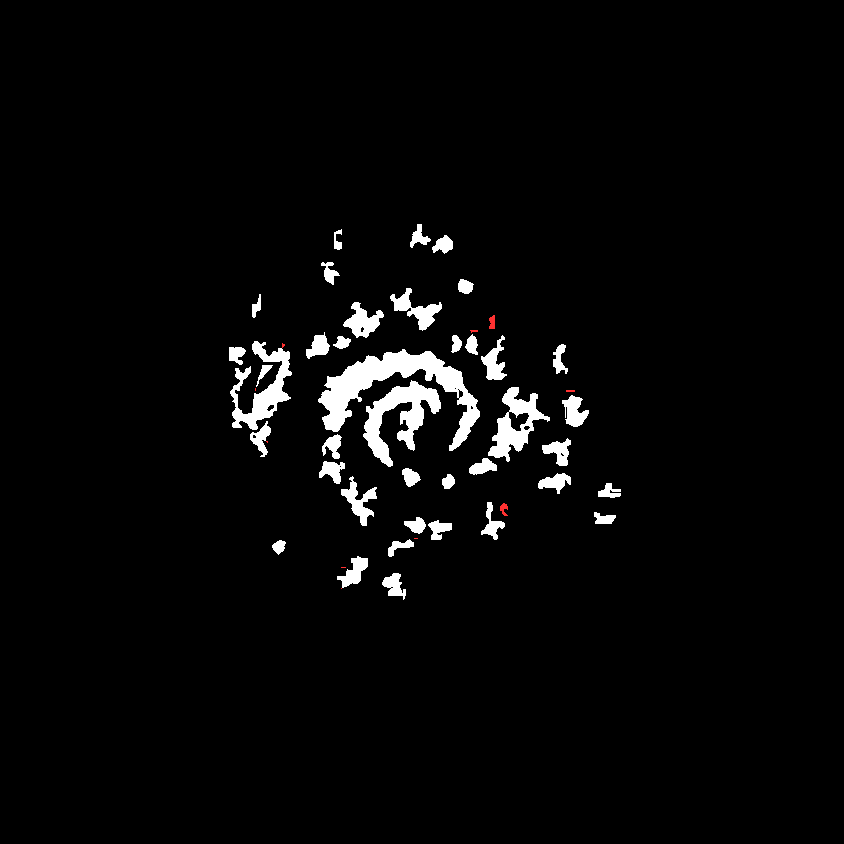}\\
  \small (g) Step 7: final cleanup (size filter repeat).
\end{minipage}\hfill
\begin{minipage}[t]{0.23\textwidth}
  \centering
  \includegraphics[width=\linewidth]{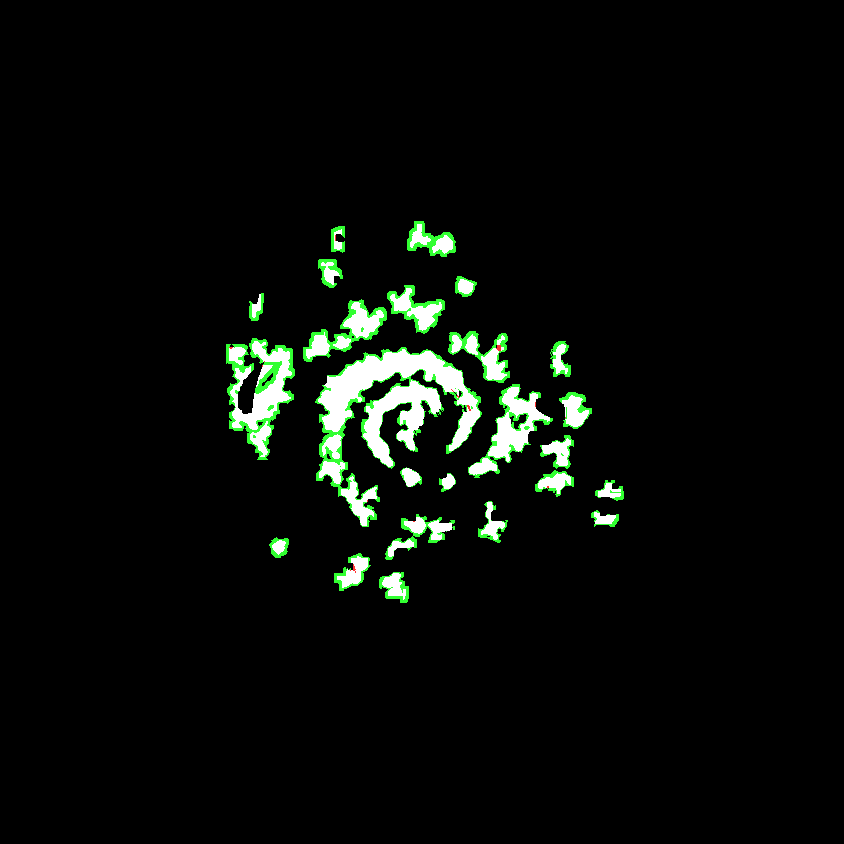}\\
  \small (h) Step 8: lost-pixel compensation.
\end{minipage}

\vspace{3mm}

\begin{minipage}[t]{0.40\textwidth}
  \centering
  \includegraphics[width=\linewidth]{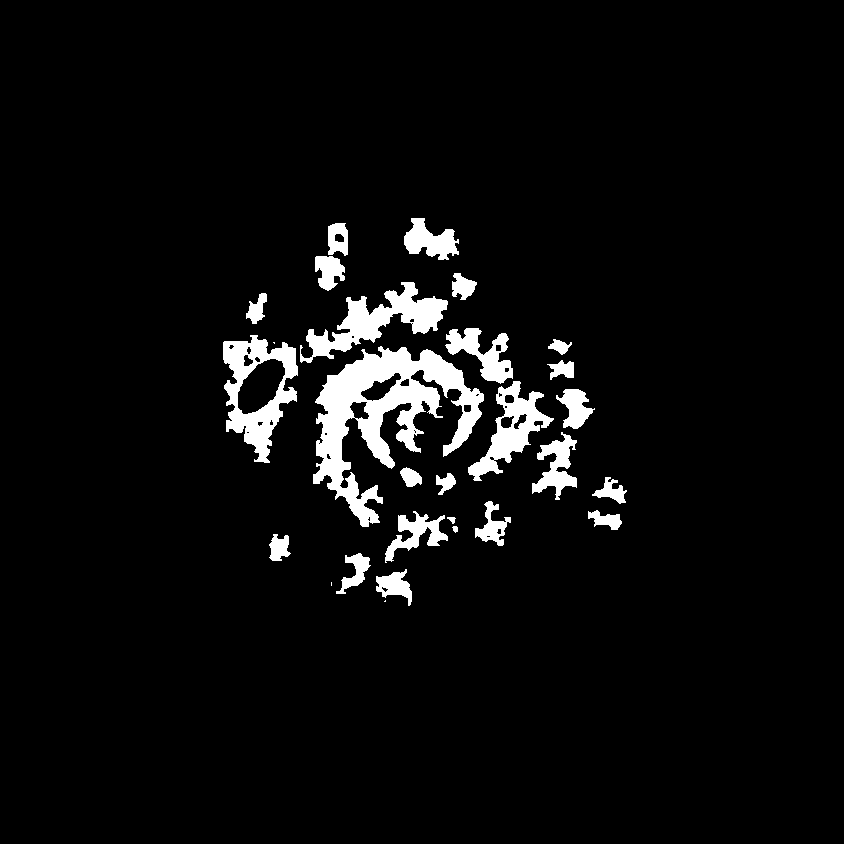}\\
  \small (i) Final refined structure mask (AMM).
\end{minipage}

\caption{Construction of the refined structure mask used to define the interarm measuring mask (IAMM).
Panels (a)–(h) show the successive processing steps applied to the iteration-0 residual detections, starting from the \textsc{SExtractor} segmentation (panel a).
Across the intermediate steps, the color coding illustrates the mask's evolution: white pixels represent the retained baseline structure, red pixels indicate artifacts or noise removed during that specific step (e.g., via morphological opening, size filtering, or consistency pruning), and green pixels highlight newly added valid signal (e.g., via gravitation or gap filling).
Panel (i) shows the final refined arm measuring mask (AMM).
The IAMM is defined as the complement of the AMM within the galaxy footprint after excluding BPM-masked pixels.}
\label{fig:appendix_mask_flow}
\end{figure*}

A visual summary of the mask-construction procedure is shown in Fig.~\ref{fig:appendix_mask_flow}. To clarify the visual representation across the panels, the color coding explicitly tracks the modifications made at each discrete step of the pipeline. In the initial detection step, the image is purely white and black, where white pixels represent the baseline structure mask directly from the \textsc{SExtractor} segmentation (constrained by the overall region mask). Next, a morphological opening ($3\times5$ elliptical kernel) is applied; here, red pixels represent tiny noise artifacts, thin bridges, and rough edges stripped away by the operation. This is followed by size filtering, where any remaining isolated structures smaller than the defined PSF area threshold are dropped from the mask, again highlighted in red. The mask then undergoes a first gravitation phase ($\max \text{distance} = 5$), with green halos indicating the mask successfully ``growing'' outward into adjacent valid, positive signal pixels. Subsequently, a smooth open/close operation ($5\times7$ kernel) refines the boundaries: red pixels indicate where sharp, protruding artifacts are eroded, while green pixels show where small gaps or bays in the structure are filled by the closing operation. A consistency constraint is then applied to anchor the mask back to the original \textsc{SExtractor} detections; any red pixels at this stage represent pruned areas that expanded during smoothing but lacked a foundation in the original detection. A final cleanup using a size filter purges the last bits of disconnected, undersized noise (red). Finally, a lost-pixel compensation step ($\max \text{distance} = 3$) performs a restricted growth phase to capture valid positive residuals immediately adjacent to the refined mask, which appear in green.

\section{Color map pixel distribution}\label{apx:color_map_pixel_distribution}

\begin{figure*}[h]
        \centering
        \includegraphics[width=0.9\hsize]{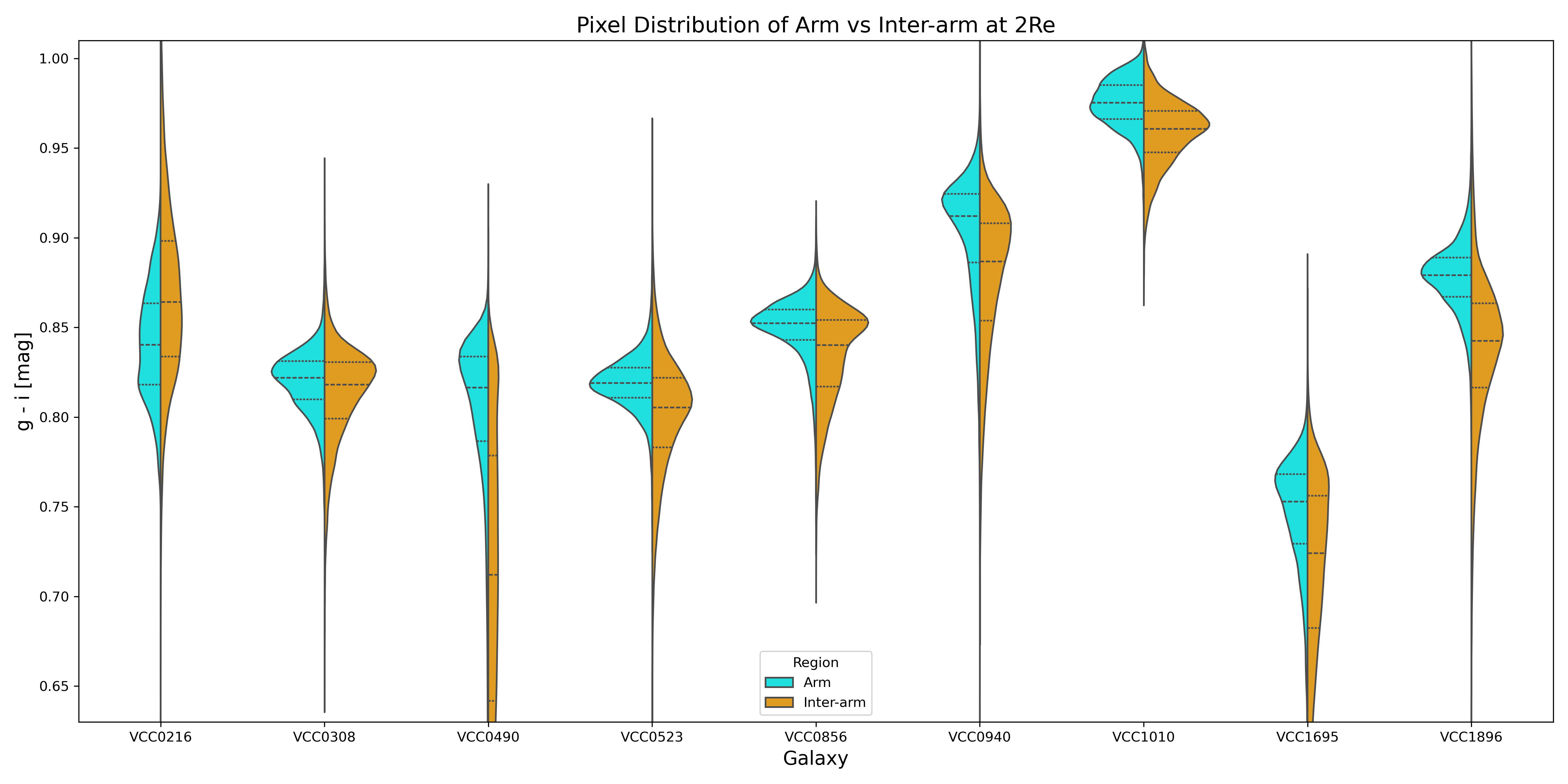}
        \caption{Pixel color distribution $g-i$ of the arm and interarm regions within $2\,R_{\mathrm{e}}$ for the nine Virgo dETGs. The data is presented as split violin plots to allow for a direct visual comparison of the probability density between the two regions. For each galaxy, the left half (cyan) shows the distribution of the arm pixels, and the right half (orange) shows the distribution of the interarm pixels. The horizontal dashed lines within each violin indicate the $25^{\text{th}}$, $50^{\text{th}}$ (median), and $75^{\text{th}}$ percentiles of the respective distributions.}
        \label{fig:color_map_distribution}%
\end{figure*}

To complement the median color measurements discussed in the main text, Fig.~\ref{fig:color_map_distribution} presents the full pixel color distributions of the arm and interarm regions within $2\,R_{\mathrm{e}}$ for the sample of nine Virgo dETGs. This figure serves as the expanded distribution version of the summarized data shown in Fig.~\ref{fig:color_map_measures}, revealing the underlying color densities, spread, and any subtle asymmetries in the $g-i$ colors that are not fully captured by the median and percentile markers alone.

\section{Residual-color map color measurement}\label{apx: residual_color_map_color_measurment}

\begin{figure*}[h]
        \centering
        \includegraphics[width=0.9\hsize]{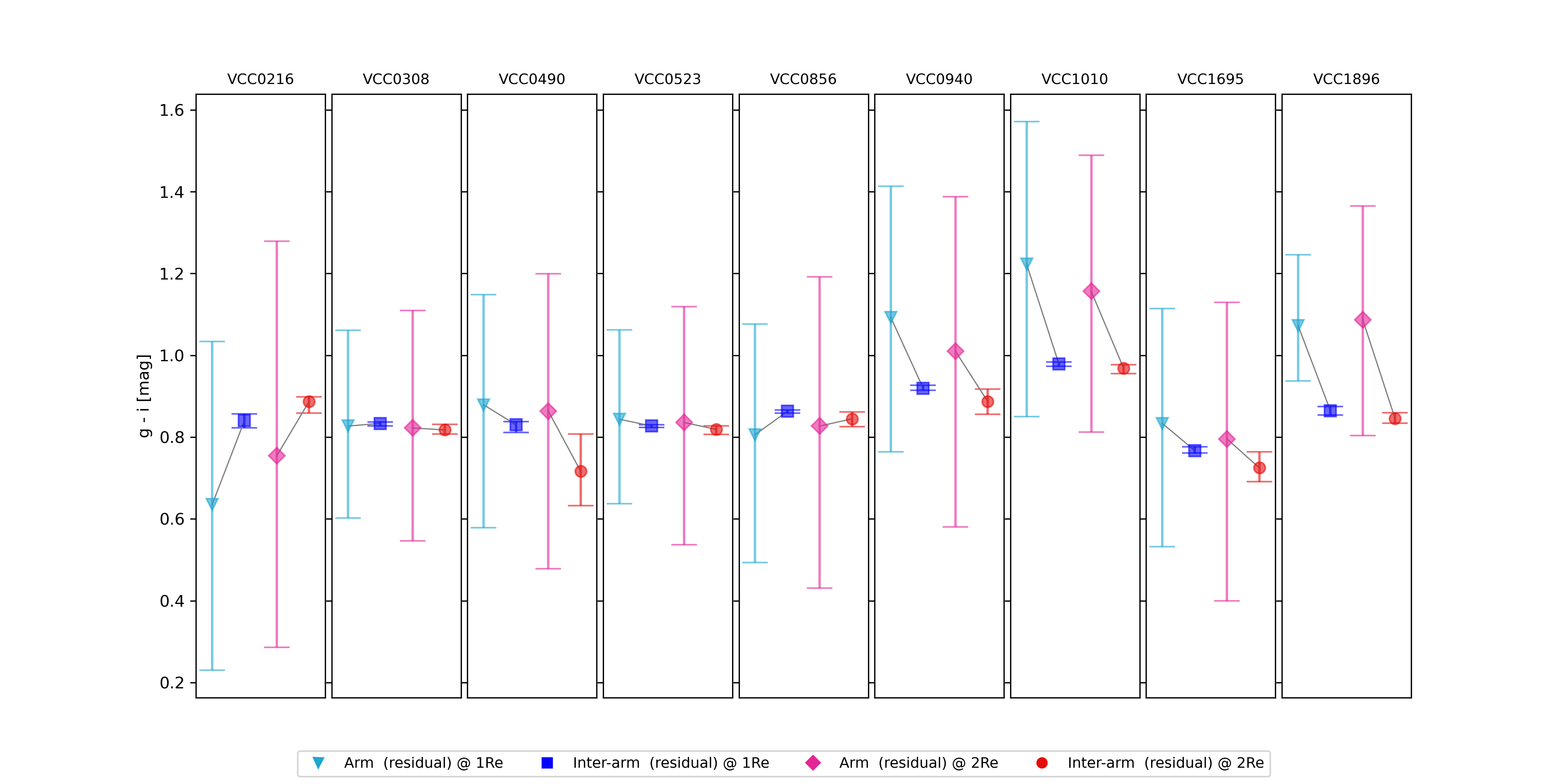}
        \caption{Color difference \(g-i\) of the interarm pixels on the diffuse-color map and arm pixels on the residual-color map image within \(1\,R_{\mathrm{e}}\) and \(2\,R_{\mathrm{e}}\) of each Virgo dwarf. The median values within \(1\,R_{\mathrm{e}}\) of the  arm and interarm   are represented by a  triangle and a square respectively, the median values within \(2\,R_{\mathrm{e}}\) of the  arm and interarm   are represented by a  diamond and a circle respectively. A gray line connects the measurements of the arm and interarm regions of a galaxy at the same effective radius. The error bars are given by the $16^{\text{th}}$ and $84^{\text{th}}$ percentiles of the distribution.}
        \label{fig:residual_color_map_measures}%
\end{figure*}

In Fig.~\ref{fig:residual_color_map_measures} we provided the residual-color map measurements for the arm and interarm regions within $1\,R_{\mathrm{e}}$ and $2\,R_{\mathrm{e}}$. The much larger percentile intervals compared with those of Fig.~\ref{fig:color_map_measures} reflect the per-pixel noise of the residual images rather than an absence of signal; as described in Sect.~\ref{sec:Results}, the residual-color maps are assessed instead through the $P(\mathrm{arm} > \mathrm{inter})$ statistic reported in Table~\ref{tab:significance_results}.

\end{appendix}

\end{document}